\documentclass[11pt]{article} 
\usepackage{fullpage}
\usepackage{graphicx}
\usepackage{upref}
\usepackage{enumerate}
\usepackage{latexsym}
\usepackage{pdfpages}
\usepackage[bookmarks]{hyperref}
\usepackage{color,graphics}
\usepackage{comment} 
\usepackage{caption}
\usepackage{subcaption}
\usepackage{wrapfig}
\usepackage{braket}

\usepackage{amssymb}
\usepackage{amsmath}
\usepackage{amsthm}
\usepackage{bm} 
\usepackage{mathrsfs}
\usepackage{mathtools}

\usepackage{epsfig,graphicx}
\usepackage{algorithmic}
\usepackage[ruled,linesnumbered]{algorithm2e}
\usepackage{amsfonts}
\usepackage{tikz} 
\usepackage{varioref}
\usepackage[ansinew]{inputenc}
\usepackage{authblk}

\theoremstyle{plain}

\theoremstyle{plain}
\newtheorem{lemma}{Lemma}[section]
\theoremstyle{plain}

\theoremstyle{plain}

\theoremstyle{plain}

\theoremstyle{plain}

\theoremstyle{definition}

\newtheorem{definition}{Definition}[section]
\theoremstyle{definition}
\newtheorem{fact}{Fact}[section]

\theoremstyle{remark}
\newtheorem{remark}{Remark}[section]

\theoremstyle{definition}

\AtBeginDocument{}

\newcommand{\fld}{\mathbb{F}}

\newcommand{\tr}{\text{Tr}}

\newcommand{\id}{\mathbb{I}}

\newcommand{\cliff}{\mathcal{C}}
\newcommand{\pauli}{\mathcal{P}}
\newcommand{\clifft}{\mathcal{J}}

\newcommand{\X}{\text{X}}
\newcommand{\Y}{\text{Y}}
\newcommand{\Z}{\text{Z}}
\newcommand{\had}{\text{H}}
\newcommand{\T}{\text{T}}
\newcommand{\CNOT}{\text{CNOT}}
\newcommand{\phase}{\text{S}}

\newcommand{\tcount}{\mathcal{T}}

\newcommand{\df}{D_F}
\newcommand{\dph}{D_P}
\newcommand{\eqpe}{\epsilon_{QPE}}
\newcommand{\eqft}{\epsilon_{QFT}}
\newcommand{\ete}{\epsilon_{TE}}
\newcommand{\bfe}{\bm{\epsilon}}

\begin{document}

\title{Composability of global phase invariant distance and its application to approximation error management}

\author[1,2]{Priyanka Mukhopadhyay \thanks{mukhopadhyay.priyanka@gmail.com, p3mukhop@uwaterloo.ca}}

\affil[1]{Institute for Quantum Computing, University of Waterloo, Canada}
\affil[2]{Department of Combinatorics and Optimization, University of Waterloo, Canada}

\maketitle

\begin{abstract}
 Many quantum algorithms can be written as a composition of unitaries, some of which can be exactly synthesized by a universal fault-tolerant gate set like Clifford+T, while others can be approximately synthesized. One task of a quantum compiler is to synthesize each approximately synthesizable unitary up to some approximation error, such that the error of the overall unitary remains bounded by a certain amount. In this paper we consider the case when the errors are measured in the global phase invariant distance. Apart from deriving a relation between this distance and the Frobenius norm, we show that this distance composes. If a unitary is written as a composition (product and tensor product) of other unitaries, we derive bounds on the error of the overall unitary as a function of the errors of the composed unitaries. Our bound is better than the sum-of-error bound, derived by Bernstein- Vazirani(1997), for the operator norm. This builds the intuition that working with the global phase invariant distance might give us a lower resource count while synthesizing quantum circuits.
 
 Next we consider the following problem. Suppose we are given a decomposition of a unitary, that is, the unitary is expressed as a composition of other unitaries. We want to distribute the errors in each component such that the resource-count (specifically T-count) is optimized. We consider the specific case when the unitary can be decomposed such that the $R_z(\theta)$ gates are the only approximately synthesizable component. We prove analytically that for both the operator norm and global phase invariant distance, the error should be distributed equally among these components (given some approximations). The optimal number of T-gates obtained by using the global phase invariant distance is less than what is obtained using the operator norm. Furthermore, we show that in case of approximate Quantum Fourier Transform, the error obtained by pruning rotation gates is less when measured in this distance, rather than the operator norm.
 
\end{abstract}



\section{Introduction}

It was envisioned \cite{1982_F,1985_D} that quantum computers can solve certain problems much more efficiently than their classical counterparts. This notion of quantum supremacy \cite{2013_P} became a possibility with the design of quantum algorithms for challenging problems like factorization \cite{1994_S,1999_S}, unstructured search \cite{1996_G} and others, with applications in areas such as cryptography \cite{2014_BB}, machine learning \cite{2019_HCTetal}, material science \cite{2020_MEABY} and quantum chemistry \cite{2017_BBSetal, 2019_LC}. 
At an abstract level, the algorithm usually consists of a number of unitaries, which are composed via tensor product or multiplication. These unitaries are mapped onto some implementation model, in most cases the circuit model. This step is a crucial part of the compilation process, where the unitary is decomposed into a number of gates (each one is a unitary) belonging to a universal set that can be implemented by the underlying technology of the hardware. The improvement claimed by the above-mentioned quantum algorithms usually do not take into account the number of resources like gates, ancilla, special states and so on, required to implement it on a particular hardware platform. The difference in the cost of implementation of these resources can make a significant difference in the practical advantage of these quantum algorithms over their classical counterparts. Thus it is necessary to estimate these resources to assess the practical advantage of quantum algorithms, to determine trade-offs (for example, problem sizes or parameters when quantum algorithms become more efficient), to determine the appropriate applications and to have a mutually reinforcing design of hardware and software.

The Clifford+T set is a widely studied and developed universal fault-tolerant gate set, which we consider in this paper. In this set the non-Clifford T gate is the most expensive to implement fault-tolerantly, as it requires large ancilla factories and additional operations like gate teleportation and state distillation \cite{1999_GC2, 2020_WWS, 2020_RBTL}, which are less accurate procedures and require additional space and time compared to a single physical gate \cite{2012_BH, 2017_CTV}. Again, not all unitaries have an exact implementation with the gates of this set. The Solovay-Kitaev algorithm \cite{1997_K, 2006_DN} guarantees that given an $n$-qubit unitary $U$, we can generate a circuit with a universal gate set like Clifford+T, such that the unitary $U'$ implemented by the circuit is at most a certain distance from $U$. A unitary $U$ is \textbf{exactly implementable} by the Clifford+T gates if there exists a circuit with these gates that implement $U$. Otherwise it is \textbf{approximately implementable} i.e. $d(U,U')\leq\epsilon$ for some $\epsilon>0$. For a distance $d$, the value $d(U,U')$ is also called the \textbf{approximation error} or error. In most synthesis and re-synthesis algorithms \cite{2020_MM, 2021_GMM, 2014_AMM, 2019_dBBW, 2015_KMM, 2016_RS, 2021_GMM2} it is required to implement a circuit for a unitary that has the minimum number of certain resource like T-gate or T-depth, or which at least reduces these resources compared to the best-known results.

Among the many kinds of approximation errors that may occur while implementing a unitary, we focus on the \textbf{synthesis errors} (see \cite{2020_MSRH} for a nice exposition). These accumulate due to the inability to implement some unitaries exactly by a discrete fault-tolerant gate set like Clifford+T. It has been proved \cite{2010_NC} that any universal fault-tolerant gate set must be discrete. In this paper our preferred distance measure (for calculating errors) is not the trace distance or operator norm used in \cite{1997_K, 2006_DN, 2013_KMM2, 2015_S, 2016_RS}, but the \textbf{global phase invariant distance} used in \cite{2015_KMM, 2011_F, 2021_GMM2} (qubit based computing), \cite{2014_KBS,2021_JS} (topological quantum computing).
This is because the trace distance or operator norm does not ignore global phase, and hence leads to unnecessarily long approximating sequences that achieve a specific global phase. In most practical applications, the global phase is not significant. In \cite{2015_KMM} the authors give an empirical formula relating the T-count of arbitrary single qubit z-rotations with approximation error $\epsilon$, where the latter is measured in the global phase invariant distance. The single-qubit z-rotation gate is defined as follows.
\begin{eqnarray}
 R_z(\theta)=\begin{bmatrix}
     e^{-i\theta/2} & 0 \\
    0 & e^{i\theta/2}
    \end{bmatrix}   \nonumber
\end{eqnarray}
In \cite{2015_S, 2016_RS} the authors give bounds on the T-count of $R_z(\theta)$ as a function of $\epsilon$, measured in the operator norm. This bound is worse than the bound obtained in \cite{2015_KMM}.

Most quantum algorithms like Quantum Fourier Transform (QFT) and phase estimation, which are also fundamental building blocks of other algorithms like factorization \cite{2010_NC}, can be written in a modular form. This implies that the overall unitary $V$ can be written as composition, i.e. multiplication and tensor product, of other unitaries, $V=\prod_{i=m}^1\bigotimes_{j=1}^{m_i}V_{ij}$. We call $V_{ij}$ as \textbf{component unitaries} in one \emph{decomposition} of $V$. There can be more than one way to decompose $V$. To reduce a particular resource of the overall unitary we would want to minimize the number of these resources in each $V_{ij}$. This can be determined by existing algorithms. For example, if the task is to optimize T-count or T-depth then we can use the T-count and T-depth optimal synthesis algorithms of \cite{2020_MM, 2021_GMM} for exactly implementable unitaries and \cite{2021_GMM2} for approximately implementable unitaries, but their complexity scales exponentially with the number of qubits. If we want a more efficient way to obtain the circuit then we can sacrifice the optimality and use re-synthesis algorithms \cite{2014_AMM, 2019_dBBW}. We can also synthesize a circuit using algorithms like \cite{1997_K, 2006_DN} and then apply a re-synthesis algorithm to reduce the T-count/depth. 

The number of resources like gate count, depth, T-count and T-depth usually are inversely proportional to the approximation error \cite{2006_DN, 2015_KMM, 2016_RS}. Given a decomposition, a compiler has to distribute the errors among each approximately-synthesizable component unitary ($V_{ij}$) such that the overall error remains bounded by some quantity. For this we need a composition rule for the distance metric in which these errors are measured. In previous works \cite{2018_HRS, 2020_MSRH} the authors worked with the operator norm and used the Bernstein-Vazirani bound \cite{1997_BV} for composing the errors. They used simulated annealing algorithm to develop an automatic error management framework that distributes the errors such that the total number of a particular resource (T-count) reduces. To the best of our knowledge, before our work, no such composition rule existed for the global phase invariant distance. 

\subsection{Our contributions}

We derive a relation between the Frobenius (and hence operator) norm and global phase invariant distance, such that an upper bound on the former implies an upper bound on the latter (Lemma \ref{lem:phFrob} in Section \ref{prelim:phFrob}). This can be useful in situations where there exists algorithms in one norm and we want to estimate some quantity in the other norm. For example, at present there exists T-count and T-depth-optimal synthesis algorithms for multi-qubit unitaries (exactly or approximately implementable) but these use global phase invariant distance \cite{2021_GMM2}. We can use these relations to get a bound on T-count or T-depth of any multi-qubit unitary in the operator norm. This is also a humble step towards understanding how the errors measured in different distances are related and compare the different algorithms. For example, we have T-count-optimal synthesis algorithms for $R_z(\theta)$ gate both in the operator norm \cite{2016_RS} as well as in the global phase invariant distance \cite{2015_KMM} and their complexities are given as function of the error in the respective distances in which they are measured. The T-count they report for the same unitary is a function of the error and it is different, higher for the operator norm. So there should be some consensus on how to compare these algorithms, for example, whether we will scale these errors and then compare the T-count or complexity or from a physical point of view there is no need for this. In order to do this, we need some relationship between the errors measured in various distances.

Apart from this, our contributions in this paper can be broadly divided into two parts.

\paragraph{1. Composition of global phase invariant distance : } First we show how the global phase invariant distance composes under multiplication and tensor product of unitaries. Let $V=\prod_{i=m}^1\bigotimes_{j=1}^{m_i}V_{ij}$ and $U=\prod_{i=m}^1\bigotimes_{j=1}^{m_i}U_{ij}$ be $n$-qubit unitaries such that $\dph(U_{ij},V_{ij})\leq\epsilon_{ij}$. We derive bounds on $\dph(U,V)$ and show this is better than the sum-of-error ($\sum_{i,j}\epsilon_{ij}$) bound.
 Since in most cases the number of resources is inversely proportional to the error, so this indicates that working in the global phase invariant distance may reduce the resource cost.
Apart from that, this result may be of independent interest, since we have a number of synthesis algorithms in qubit-based computing \cite{2015_KMM}, topological quantum computing \cite{2014_KBS, 2021_JS}, that work with global phase invariant distance, but we did not have any composition rule for this distance. 

\paragraph{2. Distribution of error while decomposing : } Next we deal with the question of how to distribute errors among the component unitaries such that we optimize the number of T-gates, while keeping the overall error bounded by some given quantity. In many popular quantum algorithms like QFT \cite{1994_S} and phase estimation \cite{1995_K} the unitary can be decomposed such that $R_z(\theta)$ gates are the only approximately synthesizable component unitaries. From \cite{2015_KMM} we know how the error relates to the T-count of $R_z(\theta)$. 

In this case, we tried to show analytically how much gain we can have by working with the global phase invariant distance, given that we have the same error bound. So we work with some approximations derived in Section \ref{compose:mult}.
This is in contrast to the approach taken in \cite{2018_HRS, 2020_MSRH} where the same problem is considered but the authors use simulated annealing to reduce the number of T-gates required, while bounding the errors by Bernstein-Vazirani bound \cite{1997_BV}. 
The bounds derived by us in Section \ref{sec:errComp} can also be used to develop a simulated annealing framework. These bounds can be used and the methods (our analytical ones as well as the automated ones in \cite{2020_MSRH}) can be adapted to reduce other resources in qubit based computing, topological quantum computing or other models where this distance is used.

 Apart from better bounds, there is another factor leading to less resources in case of approximate Quantum Fourier Transform (QFT). We show that the algorithmic error $\eqft$ accumulated due to the truncation of rotation gates \cite{2002_C}, is less if error is measured according to the global phase invariant distance. So this results in shorter circuits. QFT is an important sub-routine in other important quantum algorithms like adder, phase estimation, factoring, order finding and hidden subgroup problem \cite{2010_NC}. So a reduction in the resource count of QFT implies a resource reduction for all these algorithms.

\subsection{Related work} 

A considerable amount of work has been done to develop programming languages and toolchains for resource estimation, such as Q$\#$ \cite{2018_SGTetal}, Quipper \cite{2013_GLRSV}, Scaffold/ScaffCC \cite{2014_JPKetal}, Qiskit \cite{2019_AABetal}, staq \cite{2020_AG}, ProjectQ \cite{2018_SHT} and QuRE \cite{2013_SKFetal}. In the theoretical framework of \cite{2019_HHZetal} the authors characterize the $(Q,\lambda)$-diamond norm distance between an ideal quantum program and one which is executed on a noisy quantum hardware. This involves solving a semidefinite program \cite{2009_W} whose complexity scales exponentially with the number of qubits, making it computationally intractable for large systems. Resource estimations have also been done for specific problems such as in \cite{2017_RWSWT, 2017_SVMetal}, but usually these involve a significant amount of manual work utilizing domain knowledge.

The first designs of an automatic framework for managing approximation errors were due to Haner, Roetteler and Svore \cite{2018_HRS} and Mueli, Soeken, Roetteler and Haner \cite{2020_MSRH}. The latter improves on the former by using a fast symbolic method and incorporating two more kinds of approximation errors. Both of them use a simulated annealing program to solve an optimization problem and they work with the operator norm, one main reason being the availability of composition rules for the operator norm \cite{1997_BV}.

\subsection{Organization}

We give some preliminary definitions and results in Section \ref{sec:prelim}. The composition rules for the global phase invariant distance have been derived in Section \ref{sec:errComp}. The optimization programs have been given in Section \ref{sec:opt}. Bound on the number of T-gates for QFT and other algorithms, has been shown in Section \ref{sec:sample}. Finally we conclude in Section \ref{sec:conclude}.       

\section{Preliminaries}
\label{sec:prelim}

We denote the set of $N\times N$ $n$-qubit ($N=2^n$) unitaries by $\mathcal{U}_n$. The $(i,j)^{th}$ entry of any matrix $M$ is denoted by $M_{ij}$ or $M[i,j]$. We denote the $n\times n$ identity matrix by $\id_n$ or $\id$ if dimension is clear from the context. $[n]=\{0,1,\ldots,n-1\}$. 

\subsection{Clifford+T gate set}

The \emph{single qubit Pauli matrices} are as follows:
\begin{eqnarray}
 \X=\sigma_1=\begin{bmatrix}
     0 & 1 \\
    1 & 0
    \end{bmatrix} \qquad  
 \Y=\sigma_2=\begin{bmatrix}
     0 & -i \\
     i & 0
    \end{bmatrix} \qquad 
 \Z=\sigma_3=\begin{bmatrix}
     1 & 0 \\
     0 & -1
    \end{bmatrix}\nonumber
\label{eqn:Pauli1}
\end{eqnarray}
The \emph{$n$-qubit Pauli operators} are : $ \pauli_n=\{Q_1\otimes Q_2\otimes\ldots\otimes Q_n:Q_i\in\{\id,\X,\Y,\Z\} \}$. For convenience, we denote an $n$-qubit Pauli operator $\sigma_{a_1}\otimes\sigma_{a_2}\otimes\ldots\otimes\sigma_{a_n}$ by $\sigma_a$ where $a=(a_1,a_2,\ldots,a_n)\in [4]^n$. By convention, $\sigma_0=\id_2$.

The \emph{single-qubit Clifford group} $\cliff_1$ is generated by the Hadamard and phase gates i.e $\cliff_1=\braket{\had,\phase}$, where
\begin{eqnarray}
 \had=\frac{1}{\sqrt{2}}\begin{bmatrix}
       1 & 1 \\
       1 & -1
      \end{bmatrix}\qquad 
 \phase=\begin{bmatrix}
       1 & 0 \\
       0 & i
      \end{bmatrix} \nonumber
\end{eqnarray}
When $n>1$ the \emph{$n$-qubit Clifford group} $\cliff_n$ is generated by these two gates (acting on any of the $n$ qubits) along with the two-qubit $\CNOT=\ket{0}\bra{0}\otimes\id+\ket{1}\bra{1}\otimes\X$ gate (acting on any pair of qubits). 

The \emph{Clifford+T group} $\clifft_n$ is generated by the $n$-qubit Clifford group along with the $\T$ gate. Thus 
$\clifft_1 = \braket{\had,\T}$ and $\clifft_n=\braket{\had_{(i)},\T_{(i)},\CNOT_{(i,j)}:i,j\in [n]}$, where
\begin{eqnarray}
\T&=&\begin{bmatrix}
     1 & 0 \\
     0 & e^{i\frac{\pi}{4}}
    \end{bmatrix}\nonumber
\end{eqnarray}
A unitary $U$ is \textbf{exactly implementable} if there exists a Clifford+T circuit that implements it (up to some global phase), else it is \textbf{approximately implementable}. Specifically, we say $V$ is $\epsilon$-\textbf{approximately implementable} if there exists an exactly implementable unitary $U$ such that $d(U,V)\leq\epsilon$. We denote the set of exactly implementable unitaries by $\clifft_n$.
In this paper we use the following distance measure.
\begin{definition}[\textbf{Global phase invariant distance}]
 Given two unitaries $U,V\in\mathcal{U}_n$, we define the global phase invariant distance between them as follows.
 \begin{eqnarray}
  \dph(U,V)=\sqrt{1-\frac{\left|\tr\left(U^{\dagger}V\right)\right|}{N}}   \nonumber
 \end{eqnarray}
\end{definition}
\subsection{T-count and T-depth of circuits and unitaries}
\label{subsec:TcountDepth}

\subsubsection*{T-count and T-depth of circuits}

The \emph{T-count of a circuit} is the number of T-gates in it. 

Suppose the unitary $U$ implemented by a circuit is written as a product $U=U_mU_{m-1}\ldots U_1$ such that each $U_i$ can be implemented by a circuit in which all the gates can act in parallel or simultaneously. We say $U_i$ has depth 1 and $m$ is the \emph{depth} of the circuit. 
The \emph{T-depth of a circuit} is the number of unitaries $U_i$ where the $\T/\T^{\dagger}$ gate is the only non-Clifford gate and all the $\T/\T^{\dagger}$ gates can act in parallel. 

\subsubsection*{T-count and T-depth of exactly implementable unitaries}

The \emph{T-count of an exactly implementable unitary} $U\in\clifft_n$, denoted by $\tcount(U)$, is the minimum number of T-gates required to implement it (up to a global phase). A decomposition of $U$ with the minimum number of T-gates is called a \emph{T-count-optimal decomposition} of $U$.

The \emph{min-T-depth of an exactly 
synthesizable unitary} $U\in\clifft_n$, denoted by $\tcount_d(U)$, is the minimum T-depth of a Clifford+T circuit that implements it (up to a global phase). 
We often simply say, ``T-count'' instead of ``T-count of a unitary'' and ``T-depth'' instead of ``T-depth or min-T-depth of a unitary''. It should be clear from the context.

\subsubsection*{$\epsilon$-T-count and $\epsilon$-T-depth of approximately implementable unitaries}
\label{subsubsec:epsTcountDepth}

Let $V\in\mathcal{U}_n$ be an approximately implementable unitary. The \emph{$\epsilon$-T-count} of $V$, denoted by $\tcount^{\epsilon}(V)$, is equal to $\tcount(U)$, the T-count of an exactly implementable unitary $U\in\clifft_n$ such that $d(U,V)\leq\epsilon$ and $\tcount(U)\leq\tcount(U')$ for any $U'\in\clifft_n$ and $d(U',V)\leq\epsilon$. 

Similarly, the \emph{$\epsilon$-T-depth} of $V$, denoted by $\tcount_d^{\epsilon}(V)$, is equal to $\tcount_d(U)$, the T-depth of an exactly implementable unitary $U\in\clifft_n$ such that $d(U,V)\leq\epsilon$ and $\tcount_d(U)\leq\tcount_d(U')$ for any $U'\in\clifft_n$ and $d(U',V)\leq\epsilon$.

We call a T-count-optimal (or T-depth-optimal) circuit for any such $U$ as the \emph{$\epsilon$-T-count-optimal} (or \emph{$\epsilon$-T-depth-optimal} respectively) circuit for $V$. 

It is not hard to see that the above definitions are very general and can be applied to any unitary $V\in\mathcal{U}_n$, exactly or approximately implementable. If a unitary is exactly implementable then $\epsilon=0$. 

\subsection{Matrix norms}
\label{subsec:matNorm}

Let $\fld^{m\times n}$ be the vector space of all matrices of size $m\times n$ with entries in the field $\fld$, which is the field of real or complex numbers.

If $\|.\|$ is any norm on $\fld^n$, then the \textbf{operator norm} induced by $\|.\|$, of an $m\times n$ matrix $A$ on the space $\fld^{m\times n}$ is defined as follows.
\begin{eqnarray}
 \|A\|=\sup\left\{\|Ax\|:x\in \fld^n \text{ with } \|x\|=1 \right\} = \sup\left\{\frac{\|Ax\|}{\|x\|}:x\in \fld^n \text{ with } x\neq\vec{0} \right\}   \nonumber
 \label{eqn:opNorm}
\end{eqnarray}
In particular, the $p$-norm is defined as follows.
\begin{eqnarray}
 \|A\|_p= \sup_{x\neq\vec{0}} \frac{\|Ax\|_p}{\|x\|_p}  \nonumber
 \label{eqn:pNorm}
\end{eqnarray}
When $p=2$ then $\|A\|_2=\sigma_{\max}(A)$ where the latter is the largest singular value of $A$. This is also called the \textbf{spectral norm}.
An equivalent definition of $\|A\|_2$ is as follows.
\begin{eqnarray}
 \|A\|_2=\sup\left\{x^TAy:x,y\in \fld^n \text{ with } \|x\|_2=\|y\|_2=1  \right\}\nonumber
 \label{eqn:2norm}
\end{eqnarray}
The spectral norm is an induced or operator norm and in this paper, whenever we mention operator norm, we refer to this norm.

The \textbf{Frobenius norm} or \textbf{Hilbert-Schmidt norm} or \textbf{Schur norm} can be defined as follows.
\begin{eqnarray}
 \|A\|_F=\sqrt{\sum_{i=1}^m\sum_{j=1}^n|a_{ij}|^2}=\sqrt{\tr(A^{\dagger}A)} = \sqrt{\sum_{i=1}^{\min\{m,n\}}\sigma_i^2(A) } =\sqrt{\sum_{i=1}^R\lambda_i}\nonumber
 \label{eqn:frobNorm}
\end{eqnarray}
where $\sigma_i(A)$ are the singular values of $A$. $R\leq\min\{m,n\}$ is the rank of $A$ and $\lambda_i$ is the $i^{th}$ non-zero eigenvalue of $A^{\dagger}A$.
We also have
\begin{eqnarray}
 \|A\|_2\leq\|A\|_F \leq \sqrt{R} \|A\|_2\nonumber
 \label{eqn:2leqF}
\end{eqnarray}
\subsection{Relation between Frobenius distance and global phase invariant distance}
\label{prelim:phFrob}

Let $U$ and $V$ be two unitaries such that $\tr(V^{\dagger}U)=t$, so $\tr(U^{\dagger}V)=\overline{t}$. The global phase invariant distance between the unitaries is:
\begin{eqnarray}
 \dph(U,V) = \sqrt{1-\frac{|t|}{N}} \nonumber
\end{eqnarray}
The Frobenius distance between these unitaries is:
\begin{eqnarray}
 \df(U,V)=\|V-U\|_F=\sqrt{\tr((V-U)^{\dagger}(V-U))}    \nonumber
\end{eqnarray}
\begin{lemma}
 $$\dph(U,V)\leq \frac{\df(U,V)}{\sqrt{2N}} \qquad [N=2^n].
 $$
 \label{lem:phFrob}
\end{lemma}
\begin{proof}
For simplicity of expressions, we write $\dph$ and $\df$ instead of $\dph(U,V)$ and $\df(U,V)$ respectively.
\begin{eqnarray}
 \df^2&=& \tr(V^{\dagger}V)+\tr(U^{\dagger}U)-\tr(V^{\dagger}U)-\tr(U^{\dagger}V)   \nonumber   \\
 &=& 2N-t-\overline{t}  \qquad [\because U^{\dagger}U=V^{\dagger}V=\id]  \nonumber   \\
 &=&2N-2t_R \qquad [\text{where } t_R=Re(t)] \nonumber   \\
 \text{and }\quad N\dph^2&=&N-|t|   \nonumber
\end{eqnarray}
So
\begin{eqnarray}
 \df^2-2N\dph^2&=&2N-2t_R-2N+2|t| =2(|t|-t_R)\geq 0 \nonumber \\
 \df^2&=&2N\dph^2+2(|t|-t_R) \nonumber \\
 \implies \df^2 &\geq&2N\dph^2 \qquad [\text{Equality hold iff } |t|=t_R]  \nonumber \\
 \implies \dph&\leq& \frac{\df}{\sqrt{2N}} \qquad [\dph,\df\geq 0]  \nonumber
\end{eqnarray}
\end{proof}

The following result was proved in \cite{1997_BV}.
\begin{fact}
 If $U_1, U_1', U_2, U_2'$ are unitary transformations on an inner-product space, then 
 $$
 \|U_1'U_2'-U_1U_2\|\leq \|U_1'-U_1\|+\|U_2'-U_2\|
 $$
 \label{fact:BV97}
\end{fact}

\section{Composition of global phase invariant distance}
\label{sec:errComp}

In this section we show how the global phase invariant distance composes for tensor product and multiplication of unitaries. We prove that the bound we obtain is better than the sum-of-error bound. Ideally, it is convenient to have some compact expression that works for composition of any number of unitaries. So in Section \ref{compose:mult} we derive some approximations for multiplication of unitaries.

\subsection{Tensor product of unitaries}
\label{compose:tensor}

\begin{lemma}
 Let $V=\bigotimes_{i=m}^1V_i \in \mathcal{U}_n$, be such that $V_i\in\mathcal{U}_{n_i}$ and $\sum_{i=m}^1n_i=n$. For each $i$, $U_i$ is an $n_i$-qubit unitary such that $\dph(U_i,V_i)\leq\epsilon_i$, for some $0\leq\epsilon_i< 1$. If $U=\bigotimes_{i=m}^1U_i\in\mathcal{U}_n$, then
 $$
    \dph(U,V)\leq \sqrt{1-\prod_{i=1}^m\left(1-\epsilon_i^2\right)} 
 $$
 \label{lem:composeTensor}
\end{lemma}
\begin{proof}
 Let $N_i=2^{n_i}$ and $N=2^n=2^{\sum_{i}n_i}=\prod_{i}N_i$. Let $\tr(V_i^{\dagger}U_i)=t_i$. Since $U_i\in\mathcal{U}_n$ is such that $\dph(U_i,V_i)\leq\epsilon_i$, so we have
\begin{eqnarray}
 \dph(U_i,V_i)&=&\sqrt{1-\frac{|\tr(V_i^{\dagger}U_i)|}{N_i}}=\sqrt{1-\frac{|t_i|}{N_i}} \leq\epsilon_i  \quad [\forall i=1,\ldots,m]\nonumber \\
 \implies |t_i| &\geq& N_i\left(1-\epsilon_i^2\right). \nonumber
\end{eqnarray}
Now $\left|\tr(V^{\dagger}U)\right|=\left|\prod_{i=m}^1\tr(V_i^{\dagger}U_i)\right|=\prod_{i=m}^1|t_i|\geq \prod_{i=m}^1N_i(1-\epsilon_i^2)$, and hence
\begin{eqnarray}
\dph(U,V)&=&\sqrt{1-\frac{|\tr(V^{\dagger}U)|}{N}}\leq \sqrt{1-\prod_{i=1}^m(1-\epsilon_i^2)} \nonumber
\end{eqnarray}

\end{proof}

In the next lemma we show that the bound we obtain is better than the sum of error bound.

\begin{lemma}
 $$
    \sum_{i=1}^m\epsilon_i> \sqrt{1-\prod_{i=1}^m\left(1-\epsilon_i^2\right)} 
 $$
 when $0\leq\epsilon_i<1$.
 \label{lem:bound}
\end{lemma}

\begin{proof}
If $\sum_{i=1}^m\epsilon_i\geq 1$ then the inequality is trivially true. So we assume $\sum_{i=1}^m\epsilon_i<1$.

 It is sufficient to prove $\Delta=\left(\sum_{i=1}^m\epsilon_i\right)^2-1+\prod_{i=1}^m\left(1-\epsilon_i^2\right)>0$. We have
 \begin{eqnarray}
  \left(\sum_{i=1}^m\epsilon_i\right)^2&=&\sum_{i=1}^m\epsilon_i^2+2\sum_{i<j}\epsilon_i\epsilon_j\nonumber
 \end{eqnarray}
and
\begin{eqnarray}
\prod_{i=1}^m\left(1-\epsilon_i^2\right)&=&1-\sum_{i=1}^m\epsilon_i^2+\sum_{i<j}\epsilon_i^2\epsilon_j^2-\sum_{i<j<k}\epsilon_i^2\epsilon_j^2\epsilon_k^2+\cdots(-1)^m\prod_{i=1}^m\epsilon_i^2.\nonumber
\end{eqnarray}
So
\begin{eqnarray}
 \Delta&=&2\sum_{i<j}\epsilon_i\epsilon_j+\sum_{i<j}\epsilon_i^2\epsilon_j^2-\sum_{i<j<k}\epsilon_i^2\epsilon_j^2\epsilon_k^2+\cdots(-1)^m\prod_{i=1}^m\epsilon_i^2 \nonumber \\
 &=& 2\sum_{i<j}\epsilon_i\epsilon_j+\sum_{i<j}\epsilon_i^2\epsilon_j^2\left(1-\sum_{k=j+1}^m\epsilon_k^2\right)+\sum_{i<j<k<\ell}\epsilon_i^2\epsilon_j^2\epsilon_k^2\epsilon_{\ell}^2\left(1-\sum_{a=\ell+1}^m\epsilon_a^2\right)+\cdots  \nonumber
\end{eqnarray}
Since $\sum_{i=1}^m\epsilon_i<1$, so $\sum_{i=1}^m\epsilon_i^2<1$ and hence $\sum_{i=b}^m\epsilon_i^2<1$ for any $b\geq 1$. So each of the differences in the brackets is positive and we have $\Delta>0$. This proves the lemma.
\end{proof}

Thus in Lemma \ref{lem:composeTensor} we have $\dph(U,V)\leq\sqrt{1-\prod_{i=1}^m\left(1-\epsilon_i^2\right)}<\sum_{i=1}^m\epsilon_i$. A graphical comparison has been shown in Figure \ref{fig:tensor} of Appendix \ref{app:tensor}.

\subsection{Multiplication of unitaries}
\label{compose:mult}

\begin{lemma}
 Let $V_1,V_2,U_1,U_2\in\mathcal{U}_n$ be such that $\dph(U_1,V_1)\leq\epsilon_1$ and $\dph(U_2,V_2)\leq\epsilon_2$, where $0\leq\epsilon_1,\epsilon_2<1$. If $V=V_2V_1$ and $U=U_2U_1$ then 
 \begin{eqnarray}
  \dph(U,V)\leq\min\left\{1,\sqrt{1-\left(1-\epsilon_1^2\right)\left(1-\epsilon_2^2\right)+2\epsilon_1\epsilon_2\sqrt{\left(1-\frac{\epsilon_1^2}{2}\right)\left(1-\frac{\epsilon_2^2}{2}\right)}}\right\} <\epsilon_1+\epsilon_2  \nonumber
 \end{eqnarray}
 \label{lem:mult2}
\end{lemma}

\begin{proof}
Since $\dph(V_i,U_i)\leq\epsilon_i$ for each $i$, so from the definition of global phase invariant distance, we have
\begin{eqnarray}
 \left|\tr\left(V_i^{\dagger}U_i\right)\right|=\left|\tr\left(U_i^{\dagger}V_i\right)\right|\geq N\left(1-\epsilon_i^2\right).   \label{eqn:lemTr}
\end{eqnarray}
Let $V_2=U_2E_2$ and $V_1=E_1U_1$. Then using Inequality \ref{eqn:lemTr} we have the following.
 \begin{eqnarray}
 \left|\tr(E_1^{\dagger})\right|=\left|\tr(E_1)\right|&\geq&N(1-\epsilon_1^2)\quad\text{ and }\quad  \left|\tr(E_2^{\dagger})\right|=\left|\tr(E_2)\right|\geq N(1-\epsilon_2^2)    \label{eqn:lemTrE}
\end{eqnarray}
Both $E_1, E_2$ are unitaries, so we can expand them in the Pauli basis. Let
\begin{eqnarray}
 E_1&=&\sum_{a\in\pauli_n}e_{a1}\sigma_a\quad\text{ and }\quad E_2 =\sum_{a\in\pauli_n}e_{a2}\sigma_a.\nonumber
\end{eqnarray}
For simplicity of notations, we write $a\in\pauli_n$, instead of $\sigma_a\in\pauli_n$. From Inequality \ref{eqn:lemTrE} we can write
\begin{eqnarray}
 |e_{\id1}|&\geq& (1-\epsilon_1^2)\quad\text{ and }\quad |e_{\id2}|\geq (1-\epsilon_2^2).\nonumber
\end{eqnarray}
Since $E_1E_1^{\dagger}=E_2E_2^{\dagger}=\id$ so we have
\begin{eqnarray}
 \sum_{a\neq\id}|e_{a1}|^2&\leq& 1- (1-\epsilon_1^2)^2\leq 2\epsilon_1^2-\epsilon_1^4 \nonumber\\
 \text{and } \sum_{a\neq\id}|e_{a2}|^2&\leq& 1- (1-\epsilon_2^2)^2\leq 2\epsilon_2^2-\epsilon_2^4.\nonumber
\end{eqnarray}
Now
$
 \tr(V^{\dagger}U)=\tr(V_1^{\dagger}V_2^{\dagger}U_2U_1)=\tr((U_1V_1^{\dagger})(V_2^{\dagger}U_2))=\tr(E_1^{\dagger}E_2^{\dagger})    
$ (where the second equality follows from the invariance of trace under cyclic permutations),
and
\begin{eqnarray}
 E_1E_2&=&\left(\sum_{a\in\pauli_n}e_{a1}\sigma_a\right)\left(\sum_{a\in\pauli_n}e_{a2}\sigma_a\right)=\sum_{a}e_{a1}e_{a2}\id+\sum_{a\neq b}e_{a1}e_{b2}\sigma_a\sigma_b \nonumber \\
 &=&e_{\id1}e_{\id2}\id+\sum_{a\neq\id}e_{a1}e_{a2}\id+\sum_{a\neq\id}e_{a1}e_{\id2}\sigma_a+\sum_{b\neq\id}e_{\id1}e_{b2}\sigma_b+\sum_{a\neq b,a,b\neq\id}e_{a1}e_{a2}\sigma_a\sigma_b  \nonumber \\
 \text{So }\left|\tr(E_1^{\dagger}E_2^{\dagger})\right|= \left|\tr(E_1E_2)\right|&\geq& N(1-\epsilon_1^2)(1-\epsilon_2^2)-N\sqrt{(2\epsilon_1^2-\epsilon_1^4)(2\epsilon_2^2-\epsilon_2^4)} \nonumber 
\end{eqnarray}
and the first inequality of the lemma follows, considering the fact that the global phase invariant distance cannot be more than 1.

To prove the second inequality consider the following difference.
\begin{eqnarray}
&&1-\left(1-\epsilon_1^2\right)\left(1-\epsilon_2^2\right)+2\sqrt{\left(2\epsilon_1^2-\epsilon_1^4\right)\left(2\epsilon_2^2-\epsilon_2^4\right)} -\left(\epsilon_1+\epsilon_2\right)^2   \nonumber \\
&=&\epsilon_1^2+\epsilon_2^2-\epsilon_1^2\epsilon_2^2+2\epsilon_1\epsilon_2\sqrt{\left(1-\frac{\epsilon_1^2}{2}\right)\left(1-\frac{\epsilon_2^2}{2}\right)}-\epsilon_1^2-\epsilon_2^2-2\epsilon_1\epsilon_2    \nonumber\\
&\leq&-\epsilon_1^2\epsilon_2^2+2\epsilon_1\epsilon_2-2\epsilon_1\epsilon_2=-\epsilon_1^2\epsilon_2^2\leq 0 \nonumber
\end{eqnarray}

\end{proof}

\subsubsection*{Approximation and composition}

Let $V=\prod_{i=m}^1V_i\in\mathcal{U}_n$, be such that each $V_i\in\mathcal{U}_n$. For each $i$, $U_i$ is an $n$-qubit unitary such that $\dph(U_i,V_i)\leq\epsilon_i$, for some $0\leq\epsilon_i\leq 1$. Let $U=\prod_{i=m}^1U_i$. We want to derive a bound on $\dph(U,V)$. 

We can use the bound derived in Lemma \ref{lem:mult2} in an iterative way as follows. First, derive a bound on $\dph(U_2U_1,V_2V_1)\leq\epsilon_2'$ (Let). If $\overline{U}_2=U_2U_1$ and $\overline{V}_2=V_2V_1$, then we use Lemma \ref{lem:mult2} and derive a bound on $\dph(U_3U_2U_1,V_3V_2V_1)=\dph(U_3\overline{U}_2,V_3\overline{V}_2)$ as a function of $\epsilon_2'$ and $\epsilon_3$. Plugging in the expression for $\epsilon_2'$, we get a function of $\epsilon_1, \epsilon_2$ and $\epsilon_3$. More illustration can be found in Appendix \ref{app:graph}. Clearly, it can be hard to get a compact expression for general $m$. So we do some approximation.

\paragraph{Approximation-I : } From Lemma \ref{lem:mult2} we get the following (assuming bound is less than $1$).
\begin{eqnarray}
 \epsilon_2'^2&=&1-1+\epsilon_1^2+\epsilon_2^2-\epsilon_1^2\epsilon_2^2+2\epsilon_1\epsilon_2\sqrt{1-\frac{\epsilon_1^2}{2}-\frac{\epsilon_2^2}{2}+\frac{\epsilon_1^2\epsilon_2^2}{4}} \nonumber \\
 &\approx&\epsilon_1^2+\epsilon_2^2+2\epsilon_1\epsilon_2\sqrt{1-\epsilon_1^2-\epsilon_2^2}=\widetilde{\epsilon_2}^2\quad[\text{Let}]\qquad [\text{for small enough }\epsilon_1,\epsilon_2]   \nonumber
\end{eqnarray}
Now we use this approximate $\widetilde{\epsilon_2}$ in Lemma \ref{lem:mult2} to calculate $\dph(U_3\overline{U}_2,V_3\overline{V}_2)$. If $\widetilde{\epsilon_3}$ is the bound obtained then we can show the following.
\begin{eqnarray}
 \widetilde{\epsilon_3}^2&\approx&\sum_{i=1}^3\epsilon_i^2+2\epsilon_1\epsilon_2\sqrt{1-\epsilon_1^2-\epsilon_2^2}+2\epsilon_3(\epsilon_1+\epsilon_2)\sqrt{1-\epsilon_3^2-(\epsilon_1+\epsilon_2)^2}    \nonumber
\end{eqnarray}
Thus if we keep iterating then we can give an approximate upper bound as follows.
\begin{eqnarray}
\dph(U,V)&\leq&\min\{1,\epsilon\}   \nonumber \\
\text{where }\epsilon^2&\approx& \sum_{i=1}^m\epsilon_i^2+2\sum_{i=2}^m\epsilon_i\left(\sum_{j<i}\epsilon_j\right)\sqrt{\max\{0,1-\epsilon_i^2-(\sum_{j<i}\epsilon_j)^2\}} 
\label{eqn:approx1}
\end{eqnarray}

In Appendix \ref{app:approx1} (Figure \ref{fig:approx1}) we show that this bound closely follows the bound derived from Lemma \ref{lem:mult2} and hence can be considered as a good approximation.

\paragraph{Approximation-II : } If $\epsilon_i<0.01$ and $m$ is not large then in Lemma \ref{lem:mult2} we can ignore the summation term within the square root and write the following.
\begin{eqnarray}
 \dph(U,V)&\leq&\widehat{\epsilon_2}    \nonumber \\
 \text{where }\widehat{\epsilon_2}&\approx&c\cdot\sqrt{1-\prod_{i=1}^m(1-\epsilon_i^2)}  \label{eqn:approx2}
\end{eqnarray}
In Appendix \ref{app:approx2} (Figure \ref{fig:approx2}) we have compared this bound with the bound derived from Lemma \ref{lem:mult2} when $c=7.5$ and $m\leq 110$. For applications like resource estimation we can consider this as a good enough approximation. This bound is more compact and hence more convenient for analytical treatment. But this can be more than the bound derived from Lemma \ref{lem:mult2} when $m$ is less than $c^2\approx 56$. We can change the value of $c$ and get other approximations that work well for other range of $m$.

\begin{remark}
 From Figure \ref{fig:approx1}, \ref{fig:approx2} and \ref{fig:tensor} in Appendix \ref{app:graph} and \ref{app:tensor}, we see that the growth of error is quite less in case of tensor product of unitaries. So whenever possible, it is better to analyse by having more components in tensor or in parallel. For example, instead of treating $U=(\id\otimes U_2)\cdot (U_1\otimes\id)$, if we think $U=U_1\otimes U_2$, then the overall error is significantly less and so the estimation on resource requirement is also much less. This is another motivation for depth-optimal synthesis.
\end{remark}


\subsection{Composition of unitaries as arbitrary tensor product and multiplication}
\label{compose:arbit}

Let $V=\prod_{i=m}^1V_i$ be an approximately implementable $n$-qubit unitary which has been decomposed as multiplication of component unitaries. Each such component unitary $V_i\in\mathcal{U}_n$ is further decomposed as : $V_i=\bigotimes_{j=1}^{m_i}V_{ij}$. Let $U_{ij}$ be unitaries (of proper dimension) such that $\dph(V_{ij},U_{ij})\leq\epsilon_{ij}$, for each $i,j$. Let $U_i=\bigotimes_{j=1}^{m_i}U_{ij}$ and $U=\prod_{i=m}^1U_i$. We want to bound $\dph(V,U)$ as function of $\epsilon_{ij}$.

From Lemma \ref{lem:composeTensor} we know 
\begin{eqnarray}
 \dph(V_i,U_i)&\leq& \sqrt{1-\prod_{j=1}^{m_i}\left(1-\epsilon_{ij}^2\right)} =\epsilon_i \qquad [\forall i=1,\ldots,m] \nonumber
\end{eqnarray}
Now we can use Lemma \ref{lem:mult2} iteratively to derive $\dph(U,V)$. Alternatively, we can use the approximations. Using Equation \ref{eqn:approx1} (approximation-I) we get
\begin{eqnarray}
\dph(U,V)&\leq&\sqrt{ \sum_{i=1}^m\epsilon_i^2+2\sum_{i=2}^m\epsilon_i\left(\sum_{j<i}\epsilon_j\right)\sqrt{\max\{0,1-\epsilon_i^2-(\sum_{j<i}\epsilon_j)^2\}} } < \sum_{i=1}^m\sum_{j=1}^{m_i}\epsilon_{ij}\nonumber 
\end{eqnarray} 
and using Equation \ref{eqn:approx2} (approximation-II) we get 
\begin{eqnarray}
 \dph(V,U)&\leq& c\cdot\sqrt{1-\prod_{i=1}^m\left(1-\epsilon_i^2\right)}=c\cdot\sqrt{1-\prod_{i=1}^m\prod_{j=1}^{m_i}\left(1-\epsilon_{ij}^2\right)}
 <\sum_{i=1}^m\sum_{j=1}^{m_i}\epsilon_{ij}  .\nonumber
\end{eqnarray}

 Similarly, if we have decompositions that are tensor of product of unitaries i.e. $V=\bigotimes_{i=m}^1\prod_{j=1}^{m_i}V_{ij}$, $U=\bigotimes_{i=m}^1\prod_{j=1}^{m_i}U_{ij}$, such that $\dph(U_{ij},V_{ij})\leq\epsilon_{ij}$ then we can use the results in Section \ref{compose:tensor} and \ref{compose:mult} in order to bound $\dph(U,V)$. This is again, less than the sum-of-error bound, $\sum_{i,j}\epsilon_{ij}$. 

\section{Optimization problems}
\label{sec:opt}

The discussion in the previous section is enough to show that given the same error bound, it is advantageous to work with the global phase invariant distance, in the sense that we can allot more error per component unitary and hence get less resource estimate, which usually is inversely proportional to error. To illustrate further, in this section we formulate some optimization problems to find the distribution of approximation errors such that the total resource requirement, in our case the number of T-gates, is minimized. We solve these problems analytically for some specific scenarios and compare the resource estimates in the two distances. In case of product of unitaries we use approximation-II (Equation \ref{eqn:approx2}) and thus we assume conditions such that it is close to the exact bound derived.  

Let $C(V,\epsilon)$ be a cost function that captures the quantity we want to minimize like number of T-gates, T-depth, circuit depth and total number of gates in the circuit, for a given bound $\epsilon$ on the approximation error (in the global phase invariant distance) of the complete circuit. This means that if $U$ is the unitary implemented by the circuit and $V$ is a given unitary, then $\dph(U,V)\leq\epsilon$. Let $V=\prod_{i=m}^1\bigotimes_{j=1}^{m_i}V_{ij}$ be a decomposition of $V$ into component unitaries. $U_{ij}$ is an exactly implementable unitary such that $\dph(U_{ij},V_{ij})\leq\epsilon_{ij}$ and it requires the minimum number of resources among all unitaries that $\epsilon_{ij}$-approximates $V_{ij}$. So $U=\prod_{i=m}^1\bigotimes_{j=1}^{m_i}U_{ij}$ is the unitary implemented by the circuit. $C(U_{ij})$ is the minimum number of resources required to implement $U_{ij}$, i.e. $C(V_{\ij},\epsilon_{ij})=C(U_{ij})$. 

So our optimization program to find the minimum count of any resource is :
\begin{eqnarray}
 \min_{\bfe_{ij}} && C(V,\epsilon)=\sum_{i,j} C(V_{ij},\bfe_{ij})    \nonumber \\
 \text{s.t. } && c^2\cdot\left(1-\prod_{i=1}^m\prod_{j=1}^{m_i}\left(1-\bfe_{ij}^2\right)\right) < (\epsilon-\delta)^2 \qquad [c\text{ is a constant}]
 \label{opt:count}
\end{eqnarray}
and the program to find the minimum depth of any resource is:
\begin{eqnarray}
 \min_{\bfe_{ij}} && C(V,\epsilon)=\sum_{i} \max\{C(V_{ij},\bfe_{ij}):j=1\ldots m_i\}   \nonumber \\
 \text{s.t. } && c^2\cdot\left(1-\prod_{i=1}^m\prod_{j=1}^{m_i}\left(1-\bfe_{ij}^2\right)\right) < (\epsilon-\delta)^2
 \label{opt:depth}
\end{eqnarray}
We have reduced the upper bound in overall error in order to account for the approximation.
We have used bold fonts for variables. Since the resource count of only approximately implementable unitaries is a function of the approximation error, so 
 in the optimization programs \ref{opt:count} and \ref{opt:depth} we assume, without loss of generality, that each $V_{ij}$ is approximately synthesizable. One problem with such formulation is the number of parameters $\bfe_{ij}$, that may grow infeasibly large. In fact, the decomposition of $V$ may depend on $\epsilon$ and so the number of $V_{ij}$ (and hence $\bfe_{ij}$) may vary. A software solution was given in \cite{2018_HRS, 2020_MSRH}, where the authors used simulated annealing to solve the optimization poblem.

In this section we consider the setting where $R_z(\theta)$ gates are the only approximately synthesizable unitaries, as done in \cite{2018_HRS, 2020_MSRH}. We consider the problem of minimizing the number of T gates. Thus we use the empirical relation in \cite{2015_KMM}, by which 
\begin{eqnarray}
 C(V_{ij},\bfe_{ij})=k\log\left(\frac{1}{\bfe_{ij}}\right)+k_2\qquad \text{where } k=3.067 \text{ and } k_2=-4.322
 \label{eqn:KMM15}
\end{eqnarray}
where the error is measured in the global phase invariant distance. 
\subsection{Optimal cost of our optimization program}

We can use Karush-Kuhn-Tucker (KKT) conditions \cite{1939_K, 1951_KT, 2014_KT} to solve the above optimization problem with inequality constraint. For simplicity and without loss of much generality we use equality in the constraint and follow the Lagrangian method of optimization. Let us consider the scenario when the unitary is written as a product of $N_R$ $R_z(\theta)$ gates. 

Our optimization problem is as follows.
\begin{eqnarray}
 \min_{\bfe_i} && C(V,\epsilon)=\sum_{i=1}^{N_R}C(V_i,\bfe_i)    \nonumber \\
 \text{s.t.} && (\epsilon-\delta)^2=c^2\cdot\left(1-\prod_{i=1}^{N_R}\left(1-\bfe_i^2\right)\right) \nonumber \\
 \text{Equivalently,} && \log\left(1-\frac{(\epsilon-\delta)^2}{c^2}\right)=\sum_{i=1}^{N_R}\log\left(1-\bfe_i^2\right) \nonumber
\end{eqnarray}
Let $\lambda$ be a Lagrange multiplier. The Lagrange formulations is:
\begin{eqnarray}
  \mathcal{L}&=&\sum_{i=1}^{N_R}C(V_i,\bfe_i)+\lambda\left(\log\left(1-\frac{(\epsilon-\delta)^2}{c^2}\right)-\sum_{i=1}^{N_R}\log\left(1-\bfe_i^2\right)\right)    \nonumber
\end{eqnarray}
To optimize, the following derivatives must be zero.
\begin{eqnarray}
 \frac{\partial\mathcal{L}}{\partial\bfe_i}&=&\frac{\partial C(V_i,\bfe_i)}{\partial\bfe_i}+\lambda\frac{2\bfe_i}{1-\bfe_i^2}=0\quad\forall i=1,\ldots,N_R    \nonumber \\
 \implies 2\lambda&=&-\frac{1-\bfe_i^2}{\bfe_i}\cdot\frac{\partial C(V_i,\bfe_i)}{\partial\bfe_i}\label{eqn:lambda}
\end{eqnarray}

Since $C(V_i,\bfe_i)=k\log\left(\frac{1}{\bfe_i}\right)+k_2$ (Equation \ref{eqn:KMM15}) where $k=3.067$ and $k_2=-4.322$, we have
\begin{eqnarray}
 \frac{\partial C(V_i,\bfe_i)}{\partial\bfe_i}=-\frac{k}{\bfe_i}\nonumber
\end{eqnarray}
and Equation \ref{eqn:lambda} implies
\begin{eqnarray}
 \frac{1-\bfe_i^2}{\bfe_i}\cdot\frac{k}{\bfe_i}&=&\frac{1-\bfe_j^2}{\bfe_j}\cdot\frac{k}{\bfe_j}  \quad [i\neq j]   \nonumber \\
 \implies \bfe_i^2&=&\bfe_j^2\quad \text{Equivalently, }\bfe_i=\bfe_j=\epsilon_r\quad [\text{Let}]  \nonumber    
\end{eqnarray}
Then 
\begin{eqnarray}
  \left(1-\epsilon_r^2\right)^{N_R}&=&1-\frac{(\epsilon-\delta)^2}{c^2}   \nonumber \\
 \epsilon_r&=&\sqrt{1-\left(1-\frac{(\epsilon-\delta)^2}{c^2}\right)^{\frac{1}{N_R}}} \nonumber
\end{eqnarray}
And the optimal number of T-gates is
\begin{eqnarray}
 C(V,\epsilon)_{min}&=& = \frac{N_Rk}{2}\log\frac{1}{1-\left(1-\frac{(\epsilon-\delta)^2}{c^2}\right)^{\frac{1}{N_R}}}+N_Rk_2   \nonumber \\
 &=&\frac{3.067N_R}{2}\log\frac{1}{1-\left(1-\frac{(\epsilon-\delta)^2}{c^2}\right)^{\frac{1}{N_R}}}-4.322N_R    \label{eqn:costPh}
\end{eqnarray}

\paragraph{Tensor product :} If a unitary is written as tensor product of $R_z(\theta)$ gates then the above analysis holds with $c=1$ and $\delta=0$.

\subsubsection*{Optimal cost in \cite{2018_HRS, 2020_MSRH}}

In \cite{2018_HRS, 2020_MSRH} the authors have worked with the operator norm and used the Bernstein-Vazirani bound \cite{1997_BV}. They have considered the unitary as product of component unitaries. So the constraint function is as follows.
\begin{eqnarray}
 \sum_{i=1}^{N_R}\bfe_i&\leq&\epsilon \nonumber
\end{eqnarray}
Strictly speaking, while working with operator norm the available results \cite{2016_RS} show that the T-count of $R_z(\theta)$ vary as follows.
\begin{eqnarray}
 C(V_i,\bfe_i)&\in& 3\log\left(\frac{1}{\bfe_i}\right)+O\left(\log\log\left(\frac{1}{\bfe_i}\right)\right)
 \label{eqn:RS16}
\end{eqnarray}
There is another bound, due to Selinger \cite{2015_S}, that gives the following relation.
\begin{eqnarray}
 C(V_i,\bfe_i)&=&4\log\left(\frac{1}{\bfe_i}\right)+10
 \label{eqn:S15}
\end{eqnarray}
Clearly, both these bounds are much worse than in Equation \ref{eqn:KMM15}. Our result in Lemma \ref{lem:phFrob} show that an upper bound on operator norm implies an upper bound on global phase invariant distance. We do not know about the relation in the other direction. So the assumption that either of the above relations hold for global phase invariant distance is without rigorous mathematical proof. 

In the following we want to argue that even if we use the same bound for both the distances (the best one, Equation \ref{eqn:KMM15}), working with global phase invariant distance gives us less number of T-gates. 

For the operator norm the optimization program is as follows.
\begin{eqnarray}
 \min_{\bfe_i} && C(V,\epsilon)=\sum_{i=1}^{N_R}C(V_i,\bfe_i)   \nonumber   \\
 \text{s.t.} && \epsilon=\sum_{i=1}^{N_R}\bfe_i   
\end{eqnarray}
The Lagrangian formulation is as follows.
\begin{eqnarray}
 \mathcal{L}&=&\sum_{i=1}^{N_R}C(V_i,\bfe_i)+\lambda\left(\sum_{i=1}^{N_R}\bfe_i-\epsilon\right)\nonumber
\end{eqnarray}
To optimize, the following partial derivatives must be zero.
\begin{eqnarray}
 \frac{\partial\mathcal{L}}{\partial\bfe_i}&=&\frac{\partial C(V_i,\bfe_i)}{\partial\bfe_i}+\lambda=0\quad\forall i=1,\ldots,N_R  \nonumber \\
 \lambda&=&-\frac{\partial C(V_i,\bfe_i)}{\partial\bfe_i}=-\frac{k}{\bfe_i} \nonumber
\end{eqnarray}
Then
\begin{eqnarray}
 \frac{k}{\bfe_i}&=&\frac{k}{\bfe_j}\quad\implies\bfe_i=\bfe_j=\epsilon_r'\quad[\text{Let}].  \nonumber
\end{eqnarray}
So
\begin{eqnarray}
 N_R\epsilon_r'&=&\epsilon\quad\implies\epsilon_r'=\frac{\epsilon}{N_R}.\nonumber
\end{eqnarray}
And the optimal number of T gates is
\begin{eqnarray}
 C(V,\epsilon)_{min}&=&N_Rk\log\left(\frac{N_R}{\epsilon}\right)+N_Rk_2
 =3.067N_R\log\left(\frac{N_R}{\epsilon}\right)-4.322N_R.
 \label{eqn:costOp}
\end{eqnarray}

\subsubsection*{Comparison}

Now we calculate the difference between the optimal costs (Equations \ref{eqn:costPh} and \ref{eqn:costOp}). 
\begin{eqnarray}
 \Delta&=&N_Rk\left[\log\frac{N_R}{\epsilon}-\frac{1}{2}\log\frac{1}{1-(1-\frac{(\epsilon-\delta)^2}{c^2})^{1/N_R}}\right]=N_Rk\left[\log\sqrt{\frac{N_R^2\left(1-(1-\frac{(\epsilon-\delta)^2}{c^2})^{1/N_R}\right)}{\epsilon^2}}\right]\nonumber
\end{eqnarray}
Let $\epsilon-\delta=\epsilon'$. Then
\begin{eqnarray}
 \left(1-\frac{\epsilon'^2}{c^2}\right)^{1/N_R}&=&1-\frac{1}{N_R}\frac{\epsilon'^2}{c^2}+\frac{1}{2!}\frac{1}{N_R}(\frac{1}{N_R}-1)\frac{\epsilon'^4}{c^4}-\frac{1}{3!}\frac{1}{N_R}(\frac{1}{N_R}-1)(\frac{1}{N_R}-2)\frac{\epsilon'^6}{c^6}+\cdots\nonumber \\
 &=&1-\frac{1}{N_R}\frac{\epsilon'^2}{c^2}-\frac{N_R-1}{2!N_R^2}\frac{\epsilon'^4}{c^4}-\frac{(N_R-1)(N_R-2)}{3!N_R^3}\frac{\epsilon'^6}{c^6}-\cdots \nonumber\\
 1-\left(1-\frac{\epsilon'^2}{c^2}\right)^{1/N_R}&=&\frac{1}{N_R}\frac{\epsilon'^2}{c^2}+\frac{N_R-1}{2!N_R^2}\frac{\epsilon'^4}{c^4}+\frac{(N_R-1)(N_R-2)}{3!N_R^3}\frac{\epsilon'^6}{c^6}+\cdots   \nonumber \\
 \frac{N_R^2}{\epsilon^2}\left(1-\left(1-\frac{\epsilon'^2}{c^2}\right)^{1/N_R}\right)&=&\frac{1}{c^2}\left(1-\frac{\delta}{\epsilon}\right)^2\left[N_R+\frac{N_R-1}{2!}\frac{\epsilon'^2}{c^2}+\cdots\right] \nonumber\\
 &>&N_R\frac{1}{c^2}\left(1-\frac{\delta}{\epsilon}\right)^2\geq 1 \quad\text{ if } N_R\geq \frac{c^2}{\left(1-\frac{\delta}{\epsilon}\right)^2}\nonumber
\end{eqnarray}
For all practical purposes $1-\frac{\delta}{\epsilon}\approx 1$.
Thus $\Delta>0$ and this proves that the optimal cost i.e. number of T-gates, obtained is less if $N_R$ is roughly greater than $c^2$. We emphasize that this condition arises only because we use approximation-II. If we use the exact bound (derived from Lemma \ref{lem:mult2}) or approximation-I then we will always get less number of resources, as implied from the graphs in Appendix \ref{app:graph}.

If we do the same analysis with Equation \ref{eqn:S15} for the operator norm then we can show that error has to be distributed equally and the difference in optimal cost is as follows.
\begin{eqnarray}
 \Delta'&\approx&N_R\log\frac{N_R}{\epsilon}+14N_R+3N_R\left[\log\sqrt{\frac{N_R^2\left(1-(1-\frac{(\epsilon-\delta)^2}{c^2})^{1/N_R}\right)}{c^2}}\right]
\end{eqnarray}
which is greater than $0$ for all practical purposes.


\section{Number of T-gates for QFT and other quantum programs}
\label{sec:sample}

In this section we consider the Quantum Fourier Transform (QFT) in order to illustrate the advantage of working with global phase invariant distance. 
QFT can be used for important tasks like adder, phase estimation (QPE) and solving the hidden subgroup problem. QPE helps us to approximate the eigenvalues of a unitary operator under certain circumstances. This in turn allows us to solve other interesting problems like factoring \cite{2003_B}, order finding problem, counting solutions to a search problem. Thus a bound on the number of resources for QFT plays an important factor in the resource estimate of all these important quantum programs.
We often work with approximate QFT \cite{2002_C}, in which the number of rotations in reduced by pruning the rotation gates with small angles.

From the results in the previous sections it is clear that working with global phase invariant distance will give a lower resource count, simply because the error bound is lower than sum-of-error. For approximate QFT, there is another reason of getting lower resource estimate. We incur some error while pruning the rotation gates. We show that this error is lower if measured in the global phase invariant distance.


We consider the QFT circuit in \cite{2010_NC} that requires $N_R=\frac{n(n-1)}{2}$ controlled $R_k$ gates, where
\begin{eqnarray}
 R_k=\begin{bmatrix}
     1 & 0 \\
    0 & e^{2\pi i/2^k}
    \end{bmatrix}. \nonumber 
\end{eqnarray}
\begin{figure}[h]
\centering
\includegraphics[width=12cm, height=3cm]{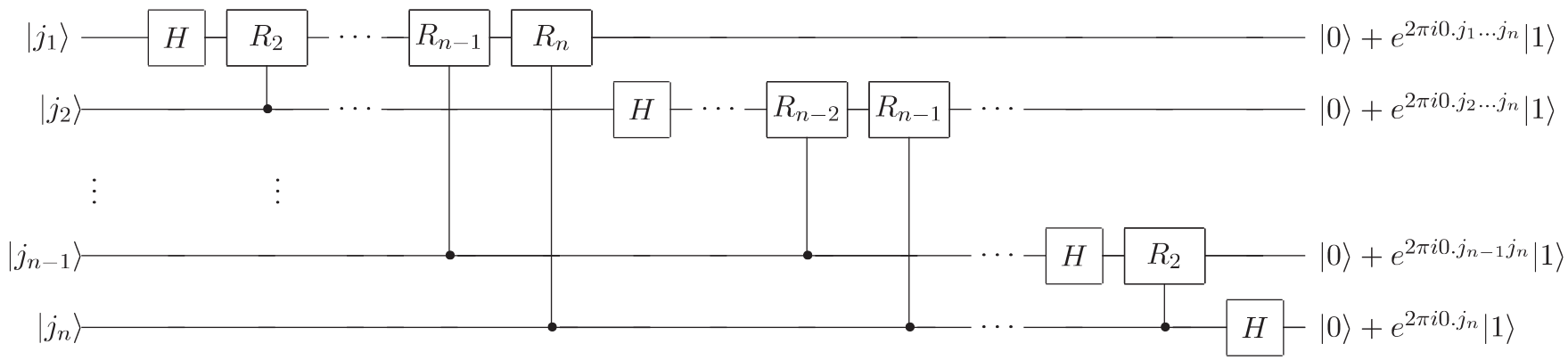}
 \caption{Quantum Fourier Transform circuit.}
 \label{fig:qft}
\end{figure}
The subscript $k$ varies from $2,\ldots,i$, when the distance between control and target qubit is $i-1$ i.e. control is at qubit 1 and $R_k$ (target) is at qubit $i$. Further decomposition into $R_z(\theta)$ gates can be obtained by two facts. First, any single qubit unitary can be decomposed in terms of two H gates and $R_z(\theta)$ gates \cite{2002_KSVV}. Second, controlled $R_z(\theta)$ can be implemented by reducing them to Fredkin and $R_z(\theta)$. Fredkin gates are exactly implementable and their T-count is 7. We focus on reducing the number of T-gates for composition of $R_z(\theta)$ gates.

In approximate QFT, pruning the rotation gates implies replacing the corresponding controlled $R_k$ gates ($cR_k$) with identity ($\id$). Then the global phase invariant distance is calculated as follows. Let $\theta_k=2\pi/2^k$.
\begin{eqnarray}
 \left|\tr\left(cR_k\id^{\dagger}\right)\right|&=& \left|3+e^{i\theta_k}\right|=\sqrt{(3+\cos \theta_k)^2+(\sin \theta_k)^2}=2\sqrt{1+3\cos^2 \frac{\theta_k}{2}} \nonumber \\
\dph(cR_k,\id)&=&\sqrt{1- \frac{\left|\tr\left(cR_k\id^{\dagger}\right)\right|}{4}}=\sqrt{1-\frac{\sqrt{1+3\cos^2 \frac{\theta_k}{2}}}{2}}=\epsilon_k' \quad [\text{Let}] \nonumber
\end{eqnarray}
Now let $\mathcal{S}=\{k:cR_k\text{ is included }\}$ and $\mathcal{S}_0=\{2,3,\ldots,n\}\setminus\mathcal{S}$. $N_k$ is the number of $cR_k$ gates in the complete circuit. 
The algorithmic error due to truncation of rotation gates, considering Equation \ref{eqn:approx1} is as follows. 
\begin{eqnarray}
\eqft=\sum_{i=1}^m\epsilon_i'^2+2\sum_{i=2}^m\epsilon_i'\left(\sum_{j<i}\epsilon_j'\right)\sqrt{\max\{0,1-\epsilon_i'^2-(\sum_{j<i}\epsilon_j')^2\}} \qquad[\text{where } m=|\mathcal{S}_0|]
\end{eqnarray}

In the operator norm we have
\begin{eqnarray}
 \|cR_k-\id\|&=&2\sin\frac{\theta_k}{2}=\widetilde{\epsilon_k'} \geq \epsilon_k'.
 \label{eqn:weqft}
\end{eqnarray}
The inequality also follows from Lemma \ref{lem:phFrob}. By Bernstein-Vazirani bound \cite{1997_BV}, here the algorithmic error due to truncation of rotation gates is $\widetilde{\eqft}\leq\sum_{k\in\mathcal{S}_0}N_k\widetilde{\epsilon_k'}$. From Inequality \ref{eqn:weqft} and our results in Section \ref{sec:errComp} we can say that $\widetilde{\eqft}>\eqft$. If we have a total error budget, then we have to allot the remaining error (i.e. excluding the approximation error due to pruning of rotation gates) among the components. Since $\widetilde{\eqft}>\eqft$, so error alloted per component in case of operator norm is less. Thus in case of approx-QFT or any unitary where such pruning occurs, there are two factors that lead to less resource count in the global phase invariant distance - (i) algorithmic error due to pruning is less, and (ii) error accumulation during composition is less. We are not giving the exact expression, since these can be either derived analytically with proper approximations, as done in Section \ref{sec:opt}, or we can use simulated annealing, as done in \cite{2018_HRS, 2020_MSRH}, with the composition rules for global phase invariant distance.

\subsection{Quantum phase estimation (QPE)}
\label{subsec:qpe}

\begin{figure}[h]
\includegraphics[width=12cm, height=4cm]{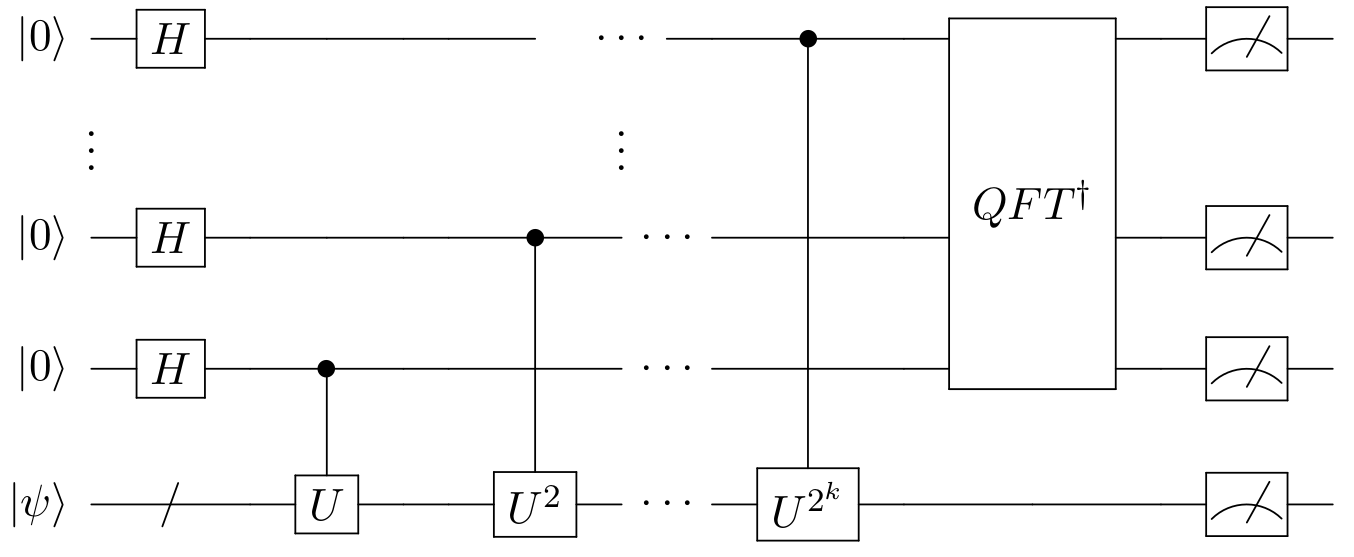}
\centering
 \caption{Quantum Phase Estimation circuit.}
 \label{fig:qpe}
\end{figure}

We have considered one of several implementations of the QPE, shown in Figure \ref{fig:qpe}. The measurement outcomes of the top $t$ qubits yield a $t$-bit approximation to the phase. $t$ also determines the probability $p$ of a successful measurement. If $n$ is the desired accuracy in number of bits, then we have the following relation \cite{2010_NC}.
\begin{eqnarray}
 t&=&n+\left\lceil\log\left(2+\frac{1}{2(1-p)}\right)\right\rceil
 \nonumber
\end{eqnarray}
Suppose $\eqpe$ is the accuracy or phase-approximation error i.e. the absolute difference between the correct phase and its $t$-bit approximation. Since this has nothing to do with the difference in the unitaries implemented, so we can assume that this error value is same for both the operator norm as well as the global phase invariant distance. In this case we can assume that we want the overall error to be bounded by $\epsilon-\eqpe$. 

Consider QPE on $U=R_z(\alpha)$. Using the results in previous sections we can conlcude that we use less number of T gates to implement the component unitaries if error is measured in global phase invariant distance, compared to the case where operator norm is used.


\section{Conclusion}
\label{sec:conclude}

In this paper we studied composability of the global phase invariant distance, which has been used for unitary synthesis in qubit based computing, topological quantum computing, etc. One of the applications of these composability rules or equations is to analyse the propagation of approximation error in circuit synthesis. This in turn helps in distribution of approximation errors during compiling such that resource cost is reduced. If a unitary is written as a composition of tensor product and multiplication of component unitaries and there are unitaries approximating each component unitary, then we show that the overall error bound is less than the sum-of-error bound, that holds for operator norm. We give approximate bounds that are more compact and convenient to work with and analyse analytically. We show how one of our approximations can be used to derive error distributions rigorously. In the special case of approximate QFT, we show we can have further advantage by analysing the error due to pruning of rotation angles in the global phase invariant distance.

 Such analysis on the resource requirements can also be done for other unitaries, for example time-evolution operators, which are widely used in simulating chemistry \cite{2017_RWSWT}. In many cases, the time-evolution operator is approximated by a Trotter-Suzuki decomposition, where the number of Trotter steps is proportional to $\frac{1}{\sqrt{\ete}}$. Here $\ete$ is the accuracy error, usually measured in the operator norm. It will be interesting to probe if the global phase invariant distance has some operational meaning here and if so, whether working in this distance gives us any advantage, for example in terms of gate count. Further applications of the results derived in this paper, have been left for future work.

\section*{Acknowledgement}  
The author would like to thank Michele Mosca for pointing out the papers \cite{2018_HRS, 2020_MSRH}, that has led to this work, and his suggestions on the initial manuscript. The author wishes to thank NTT Research for their financial and technical support. Research at IQC is supported in part by the Government of Canada through Innovation, Science and Economic Development Canada. The author would like to thank the anonymous reviewers for helpful comments.

\bibliographystyle{alpha}

\begin{thebibliography}{MEAG{\etalchar{+}}20}

\bibitem{1982_F}
Richard~P Feynman.
\newblock Simulating physics with computers.
\newblock {\em Int. J. Theor. Phys}, 21(6/7), 1982.

\bibitem{1985_D}
David Deutsch.
\newblock Quantum theory, the Church--Turing principle and the universal
  quantum computer.
\newblock {\em Proceedings of the Royal Society of London. A. Mathematical and
  Physical Sciences}, 400(1818):97--117, 1985.

\bibitem{2013_P}
John Preskill.
\newblock Quantum computing and the entanglement frontier.
\newblock {\em Bulletin of the American Physical Society}, 58, 2013.  

\bibitem{1994_S}
Peter~W Shor.
\newblock Algorithms for quantum computation: Discrete logarithms and factoring.
\newblock In {\em Proceedings 35th annual symposium on foundations of computer
  science}, pages 124--134. IEEE, 1994.

\bibitem{1999_S}
Peter~W Shor.
\newblock Polynomial-time algorithms for prime factorization and discrete logarithms on a quantum computer.
\newblock {\em SIAM review}, 41(2):303--332, 1999.
 
\bibitem{1996_G}
Lov~K Grover.
\newblock A fast quantum mechanical algorithm for database search.
\newblock In {\em Proceedings of the twenty-eighth annual ACM symposium on Theory of computing}, pages 212--219, 1996.

\bibitem{2014_BB}
Charles~H Bennett and Gilles Brassard.
\newblock Quantum cryptography: Public key distribution and coin tossing.
\newblock {\em Theoretical Computer Science}, 560:7--11, 2014.

\bibitem{2019_HCTetal}
Vojt{\v{e}}ch Havl{\'\i}{\v{c}}ek, Antonio~D C{\'o}rcoles, Kristan Temme,
  Aram~W Harrow, Abhinav Kandala, Jerry~M Chow, and Jay~M Gambetta.
\newblock Supervised learning with quantum-enhanced feature spaces.
\newblock {\em Nature}, 567(7747):209--212, 2019.

\bibitem{2020_MEABY}
Sam McArdle, Suguru Endo, Alan Aspuru-Guzik, Simon~C Benjamin, and Xiao Yuan.
\newblock Quantum computational chemistry.
\newblock {\em Reviews of Modern Physics}, 92(1):015003, 2020.

\bibitem{2017_BBSetal}
Ryan Babbush, Dominic~W Berry, Yuval~R Sanders, Ian~D Kivlichan, Artur Scherer,
  Annie~Y Wei, Peter~J Love, and Al{\'a}n Aspuru-Guzik.
\newblock Exponentially more precise quantum simulation of fermions in the
  configuration interaction representation.
\newblock {\em Quantum Science and Technology}, 3(1):015006, 2017.

\bibitem{2019_LC}
Guang~Hao Low and Isaac~L Chuang.
\newblock Hamiltonian simulation by qubitization.
\newblock {\em Quantum}, 3:163, 2019.

\bibitem{1999_GC2}
Daniel Gottesman and Isaac~L Chuang.
\newblock Demonstrating the viability of universal quantum computation using
  teleportation and single-qubit operations.
\newblock {\em Nature}, 402(6760):390--393, 1999.

\bibitem{2020_WWS}
Xin Wang, Mark~M Wilde, and Yuan Su.
\newblock Efficiently computable bounds for magic state distillation.
\newblock {\em Physical Review Letters}, 124(9):090505, 2020.

\bibitem{2020_RBTL}
Bartosz Regula, Kaifeng Bu, Ryuji Takagi, and Zi-Wen Liu.
\newblock Benchmarking one-shot distillation in general quantum resource theories.
\newblock {\em Physical Review A}, 101(6):062315, 2020.

\bibitem{2012_BH}
Sergey Bravyi and Jeongwan Haah.
\newblock Magic-state distillation with low overhead.
\newblock {\em Physical Review A}, 86(5):052329, 2012.

\bibitem{2017_CTV}
Earl~T Campbell, Barbara~M Terhal, and Christophe Vuillot.
\newblock Roads towards fault-tolerant universal quantum computation.
\newblock {\em Nature}, 549(7671):172--179, 2017.

\bibitem{1997_K}
Aleksei~Yur'evich Kitaev.
\newblock Quantum computations: algorithms and error correction.
\newblock {\em Uspekhi Matematicheskikh Nauk}, 52(6):53--112, 1997.

\bibitem{2006_DN}
CM~Dawson and MA~Nielsen.
\newblock The Solovay-Kitaev algorithm.
\newblock {\em Quantum Information and Computation}, 6(1):81--95, 2006.

\bibitem{2020_MM}
Michele Mosca and Priyanka Mukhopadhyay.
\newblock A polynomial time and space heuristic algorithm for T-count.
\newblock {\em Quantum Science and Technology}, 7(1):015003, 2021.

\bibitem{2021_GMM}
Vlad Gheorghiu, Michele Mosca, and Priyanka Mukhopadhyay.
\newblock A (quasi-)polynomial time heuristic algorithm for synthesizing T-depth
  optimal circuits.
\newblock {\em arXiv preprint arXiv:2101.03142}, 2021.

\bibitem{2014_AMM}
Matthew Amy, Dmitri Maslov, and Michele Mosca.
\newblock Polynomial-time T-depth optimization of clifford+T circuits via
  matroid partitioning.
\newblock {\em IEEE Transactions on Computer-Aided Design of Integrated
  Circuits and Systems}, 33(10):1476--1489, 2014.
  
\bibitem{2019_dBBW}
Niel de~Beaudrap, Xiaoning Bian, and Quanlong Wang.
\newblock Techniques to reduce $\pi/4$-parity phase circuits, motivated by the
  ZX calculus.
\newblock {\em arXiv preprint arXiv:1911.09039}, 2019.  


\bibitem{2015_KMM}
Vadym Kliuchnikov, Dmitri Maslov, and Michele Mosca.
\newblock Practical approximation of single-qubit unitaries by single-qubit
  quantum clifford and T circuits.
\newblock {\em IEEE Transactions on Computers}, 65(1):161--172, 2015.

\bibitem{2016_RS}
Neil~J Ross and Peter Selinger.
\newblock Optimal ancilla-free clifford+T approximation of Z-rotations.
\newblock {\em Quantum Information \& Computation}, 16(11-12):901--953, 2016.

\bibitem{2021_GMM2}
Vlad Gheorghiu, Michele Mosca, and Priyanka Mukhopadhyay.
\newblock T-count and T-depth of \emph{any} multi-qubit unitary.
\newblock {\em arXiv preprint arXiv:2110:10292}, 2021.

\bibitem{2020_MSRH}
Giulia Meuli, Mathias Soeken, Martin Roetteler, and Thomas H{\"a}ner.
\newblock Enabling accuracy-aware quantum compilers using symbolic resource
  estimation.
\newblock {\em Proceedings of the ACM on Programming Languages},
  4(OOPSLA):1--26, 2020.

\bibitem{2010_NC}
Michael~A Nielsen and Isaac~L Chuang.
\newblock {\em Quantum Computation and Quantum Information}.
\newblock Cambridge University Press, 2010.  

\bibitem{2013_KMM2}
Vadym Kliuchnikov, Dmitri Maslov, and Michele Mosca.
\newblock Asymptotically optimal approximation of single qubit unitaries by
  clifford and T circuits using a constant number of ancillary qubits.
\newblock {\em Physical Review Letters}, 110(19):190502, 2013.

\bibitem{2015_S}
Peter Selinger.
\newblock Efficient clifford+T approximation of single-qubit operators.
\newblock {\em Quantum Information \& Computation}, 15(1-2):159--180, 2015.

\bibitem{2011_F}
Austin~G Fowler.
\newblock Constructing arbitrary Steane code single logical qubit
  fault-tolerant gates.
\newblock {\em Quantum Information \& Computation}, 11(9-10):867--873, 2011.

\bibitem{2014_KBS}
Vadym Kliuchnikov, Alex Bocharov, and Krysta~M Svore.
\newblock Asymptotically optimal topological quantum compiling.
\newblock {\em Physical Review Letters}, 112(14):140504, 2014.

\bibitem{2021_JS}
Emil~G{\'e}netay Johansen and Tapio Simula.
\newblock Fibonacci anyons versus majorana fermions: A Monte Carlo approach to
  the compilation of braid circuits in $SU(2)_k$ anyon models.
\newblock {\em PRX Quantum}, 2(1):010334, 2021.

\bibitem{2002_KSVV}
Alexei~Yu Kitaev, Alexander Shen, Mikhail~N Vyalyi, and Mikhail~N Vyalyi.
\newblock {\em Classical and quantum computation}.
\newblock Number~47. American Mathematical Soc., 2002.

\bibitem{2018_HRS}
Thomas H{\"a}ner, Martin Roetteler, and Krysta~M Svore.
\newblock Managing approximation errors in quantum programs.
\newblock {\em arXiv preprint arXiv:1807.02336}, 2018.

\bibitem{1997_BV}
Ethan Bernstein and Umesh Vazirani.
\newblock Quantum complexity theory.
\newblock {\em SIAM Journal on computing}, 26(5):1411--1473, 1997.

\bibitem{1995_K}
A~Yu Kitaev.
\newblock Quantum measurements and the abelian stabilizer problem.
\newblock {\em arXiv preprint quant-ph/9511026}, 1995.

\bibitem{2002_C}
Don Coppersmith.
\newblock An approximate fourier transform useful in quantum factoring.
\newblock {\em arXiv preprint quant-ph/0201067}, 2002.

\bibitem{2018_SGTetal}
Krysta Svore, Alan Geller, Matthias Troyer, John Azariah, Christopher Granade,
  Bettina Heim, Vadym Kliuchnikov, Mariia Mykhailova, Andres Paz, and Martin
  Roetteler.
\newblock Q\# enabling scalable quantum computing and development with a
  high-level dsl.
\newblock In {\em Proceedings of the Real World Domain Specific Languages
  Workshop 2018}, pages 1--10, 2018.

\bibitem{2013_GLRSV}
Alexander~S Green, Peter~LeFanu Lumsdaine, Neil~J Ross, Peter Selinger, and
  Beno{\^\i}t Valiron.
\newblock Quipper: a scalable quantum programming language.
\newblock In {\em Proceedings of the 34th ACM SIGPLAN conference on Programming
 language design and implementation}, pages 333--342, 2013.  
  
\bibitem{2014_JPKetal}
Ali JavadiAbhari, Shruti Patil, Daniel Kudrow, Jeff Heckey, Alexey Lvov,
  Frederic~T Chong, and Margaret Martonosi.
\newblock Scaffcc: A framework for compilation and analysis of quantum
  computing programs.
\newblock In {\em Proceedings of the 11th ACM Conference on Computing
  Frontiers}, pages 1--10, 2014.
  
\bibitem{2019_AABetal}
Gadi Aleksandrowicz, Thomas Alexander, Panagiotis Barkoutsos, Luciano Bello,
  Yael Ben-Haim, David Bucher, Francisco~Jose Cabrera-Hern{\'a}ndez, Jorge
  Carballo-Franquis, Adrian Chen, Chun-Fu Chen, et~al.
\newblock Qiskit: An open-source framework for quantum computing.
\newblock {\em Accessed on: Mar}, 16, 2019.

\bibitem{2020_AG}
Matthew Amy and Vlad Gheorghiu.
\newblock STAQ - a full-stack quantum processing toolkit.
\newblock {\em Quantum Science and Technology}, 5(3):034016, 2020.

\bibitem{2018_SHT}
Damian~S Steiger, Thomas H{\"a}ner, and Matthias Troyer.
\newblock ProjectQ: an open source software framework for quantum computing.
\newblock {\em Quantum}, 2:49, 2018.

\bibitem{2013_SKFetal}
Martin Suchara, John Kubiatowicz, Arvin Faruque, Frederic~T Chong, Ching-Yi
  Lai, and Gerardo Paz.
\newblock Qure: The quantum resource estimator toolbox.
\newblock In {\em 2013 IEEE 31st International Conference on Computer Design
  (ICCD)}, pages 419--426. IEEE, 2013.

\bibitem{2019_HHZetal}
Shih-Han Hung, Kesha Hietala, Shaopeng Zhu, Mingsheng Ying, Michael Hicks, and
  Xiaodi Wu.
\newblock Quantitative robustness analysis of quantum programs.
\newblock {\em Proceedings of the ACM on Programming Languages}, 3(POPL):1--29,
  2019.  
  
\bibitem{2009_W}
John Watrous.
\newblock Semidefinite programs for completely bounded norms.
\newblock {\em Theory OF Computing}, 5:217--238, 2009.  

\bibitem{2017_RWSWT}
Markus Reiher, Nathan Wiebe, Krysta~M Svore, Dave Wecker, and Matthias Troyer.
\newblock Elucidating reaction mechanisms on quantum computers.
\newblock {\em Proceedings of the National Academy of Sciences},
  114(29):7555--7560, 2017.

\bibitem{2017_SVMetal}
Artur Scherer, Beno{\^\i}t Valiron, Siun-Chuon Mau, Scott Alexander, Eric
  Van~den Berg, and Thomas~E Chapuran.
\newblock Concrete resource analysis of the quantum linear-system algorithm
  used to compute the electromagnetic scattering cross section of a 2D target.
\newblock {\em Quantum Information Processing}, 16(3):1--65, 2017.

\bibitem{1939_K}
William Karush.
\newblock Minima of functions of several variables with inequalities as side
  constraints.
\newblock {\em M.Sc. Dissertation. Dept. of Mathematics, Univ. of Chicago},
  1939.

\bibitem{1951_KT}
HW~Kuhn, AW~Tucker, et~al.
\newblock Nonlinear programming.
\newblock In {\em Proceedings of the Second Berkeley Symposium on Mathematical
  Statistics and Probability}. The Regents of the University of California,
  1951.

\bibitem{2014_KT}
Harold~W Kuhn and Albert~W Tucker.
\newblock Nonlinear programming.
\newblock In {\em Traces and emergence of nonlinear programming}, pages
  247--258. Springer, 2014.

\bibitem{2003_B}
Stephane Beauregard.
\newblock Circuit for Shor's algorithm using 2n+3 qubits.
\newblock {\em Quantum Information \& Computation}, 3(2):175--185, 2003.


\end{thebibliography}
\newcommand{\etalchar}[1]{$^{#1}$}

\appendix 
\section{Comparison of bounds derived in Section \ref{compose:mult}}
\label{app:graph}

Let us consider unitaries $V=\prod_{i=m}^1V_i$ and $U=\prod_{i=m}^1U_i$, such that $\dph(U_i,V_i)\leq\epsilon_i$. Our aim is to show how error propagates, considering the bounds given in Section \ref{compose:mult} - with and without the approximations. It will also show how close our approximation is. 
As per convention, we assume execution of the circuit is from right to left, i.e first $U_1$ then $U_2$ and so on. We consider the distance between the unitaries after implementation of each $U_i$ i.e. $\dph(U_1,V_1)$, $\dph(U_2U_1,V_2V_1)$, $\dph(U_3U_2U_1,V_3V_2V_1)$ and so on. Equivalently, we calculate $\dph(\overline{U}_m,\overline{V}_m)$, where $\overline{U}_m=\prod_{i=m}^1U_i$, $\overline{V}_m=\prod_{i=m}^1V_i$ and $m=1,2,3,\ldots$. In our experiment, we consider $\epsilon_i=\epsilon_j$ for each $i, j$. 

\subsection{Comparison with Approximation-I}
\label{app:approx1}

First, we compare $\dph(\overline{U}_m,\overline{V}_m)$, the bound derived from Lemma \ref{lem:mult2}, Equation \ref{eqn:approx1} and the sum-of-error bound. It is straightforward to use Equation \ref{eqn:approx1} for different values of $m$. We show briefly how to use Lemma \ref{lem:mult2} to calculate the distance without approximation. 

Suppose we take $\epsilon_i=0.01$ for each $i$. So taking $m=1$ we have $\dph(U_1,V_1)\leq 0.01$. When $m=2$ then from Lemma \ref{lem:mult2} we have 
\begin{eqnarray}
 \dph(U_2U_1,V_2V_1)\leq\sqrt{1-(1-\epsilon_1^2)(1-\epsilon_2^2)+\sqrt{(2\epsilon_1^2-\epsilon_1^4)(2\epsilon_2^2-\epsilon_2^4)}}=0.019999=\widetilde{\epsilon_2}\quad [\text{Let}] \nonumber
\end{eqnarray}
and from Equation \ref{eqn:approx1} we have
\begin{eqnarray}
 \dph'(U_2U_1,V_2V_1)\leq\sqrt{\epsilon_1^2+\epsilon_2^2+2\epsilon_2\epsilon_1\sqrt{1-\epsilon_2^2-\epsilon_1^2}}=0.019992. \nonumber
\end{eqnarray}
We denote the distance calculated using approximation-I (Equation \ref{eqn:approx1}) by $\dph'(.,.)$.
When $m=3$ then using Lemma \ref{lem:mult2} we get
\begin{eqnarray}
 \dph(U_3U_2U_1,V_3V_2V_1)\leq\sqrt{1-(1-\widetilde{\epsilon_2}^2)(1-\epsilon_3^2)+\sqrt{(2\widetilde{\epsilon_2}^2-\widetilde{\epsilon_2}^4)(2\epsilon_3^2-\epsilon_3^4)}}=0.029981=\widetilde{\epsilon_3}\quad [\text{Let}]   \nonumber
\end{eqnarray}
and from Equation \ref{eqn:approx1} we have
\begin{eqnarray}
 \dph'(U_3U_2U_1,V_3V_2V_1)&\leq&\sqrt{\epsilon_1^2+\epsilon_2^2+\epsilon_3^2+2\epsilon_2\epsilon_1\sqrt{1-\epsilon_2^2-\epsilon_1^2}+2\epsilon_3(\epsilon_1+\epsilon_2)\sqrt{1-\epsilon_3^2-(\epsilon_1+\epsilon_2)^2}} \nonumber\\
 &=&0.02998 \nonumber
\end{eqnarray}
We can keep calculating the distance values for different values of $m$. This will show the error propagation or error accumulation in the overall unitary i.e. distance between $\prod_{i}U_i$ and $\prod_iV_i$.
\begin{figure}
 \centering
 \begin{subfigure}[b]{0.475\textwidth}
 \centering
  \includegraphics[width=7cm,height=5cm]{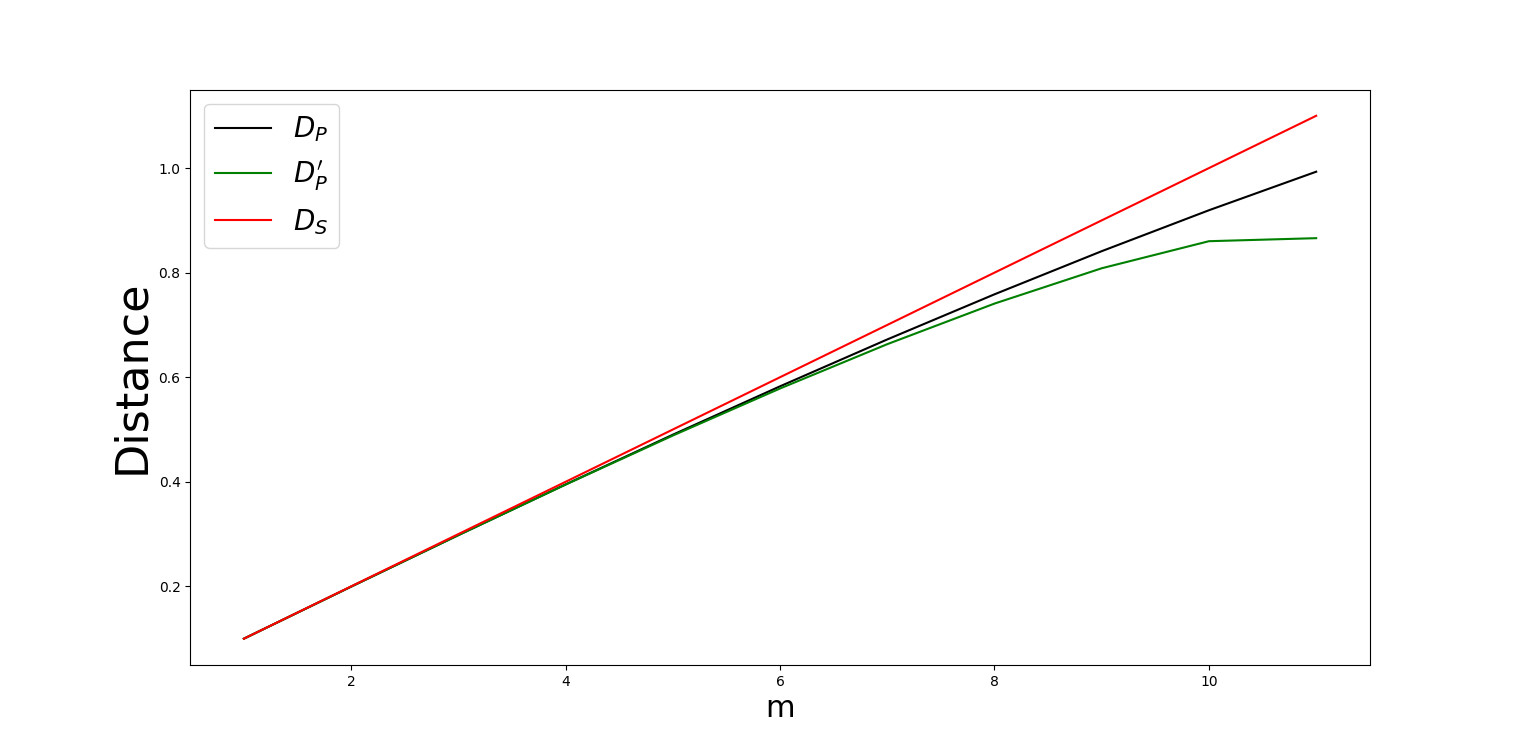}
  \caption{$\epsilon=0.1, m\leq 11$}
  \label{fig:A}
 \end{subfigure}
 \hfill
 \begin{subfigure}[b]{0.475\textwidth}
 \centering
 \includegraphics[width=7cm,height=5cm]{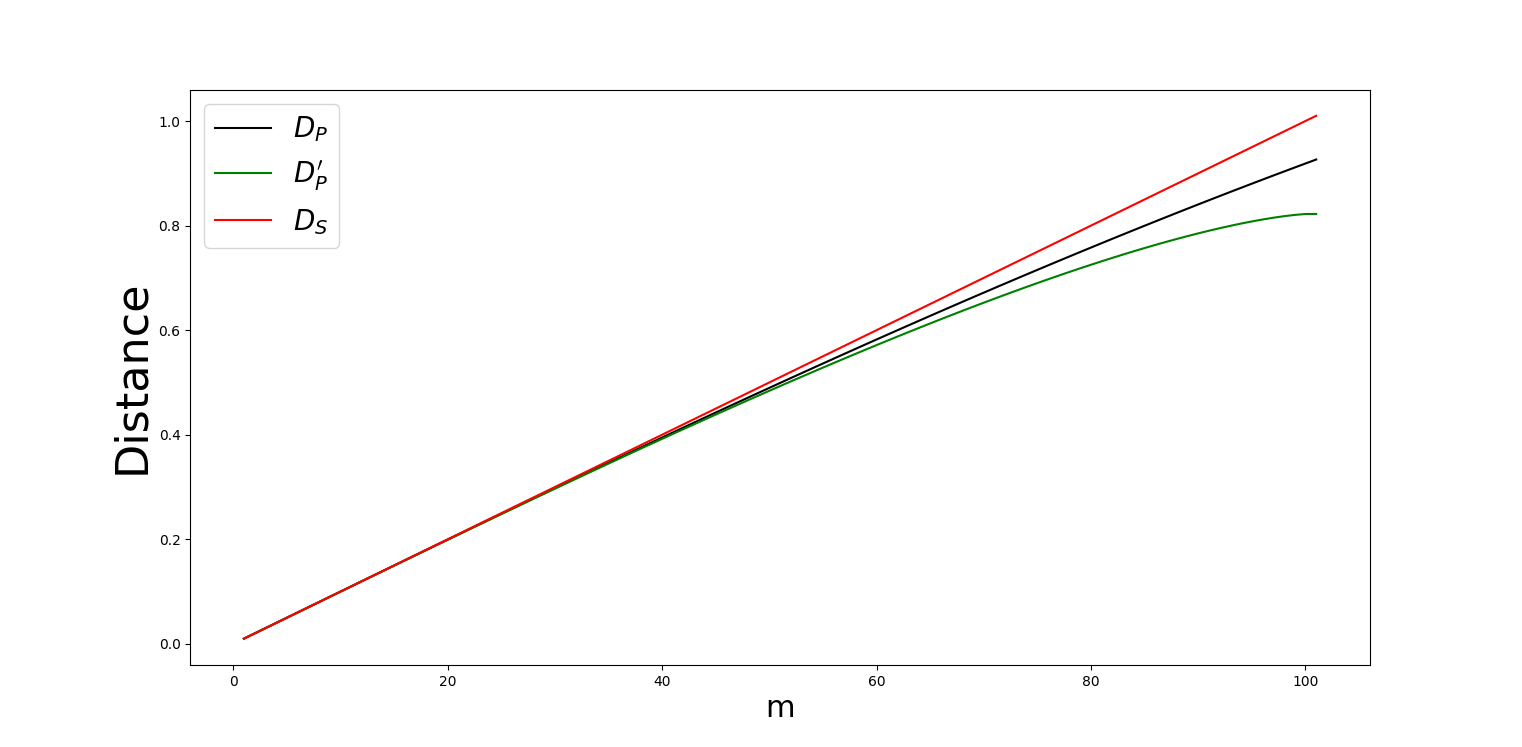}
  \caption{$\epsilon=0.01, m\leq 101$}
  \label{fig:B}
 \end{subfigure}
 \vskip\baselineskip
 \begin{subfigure}[b]{0.475\textwidth}
 \centering
 \includegraphics[width=7cm,height=5cm]{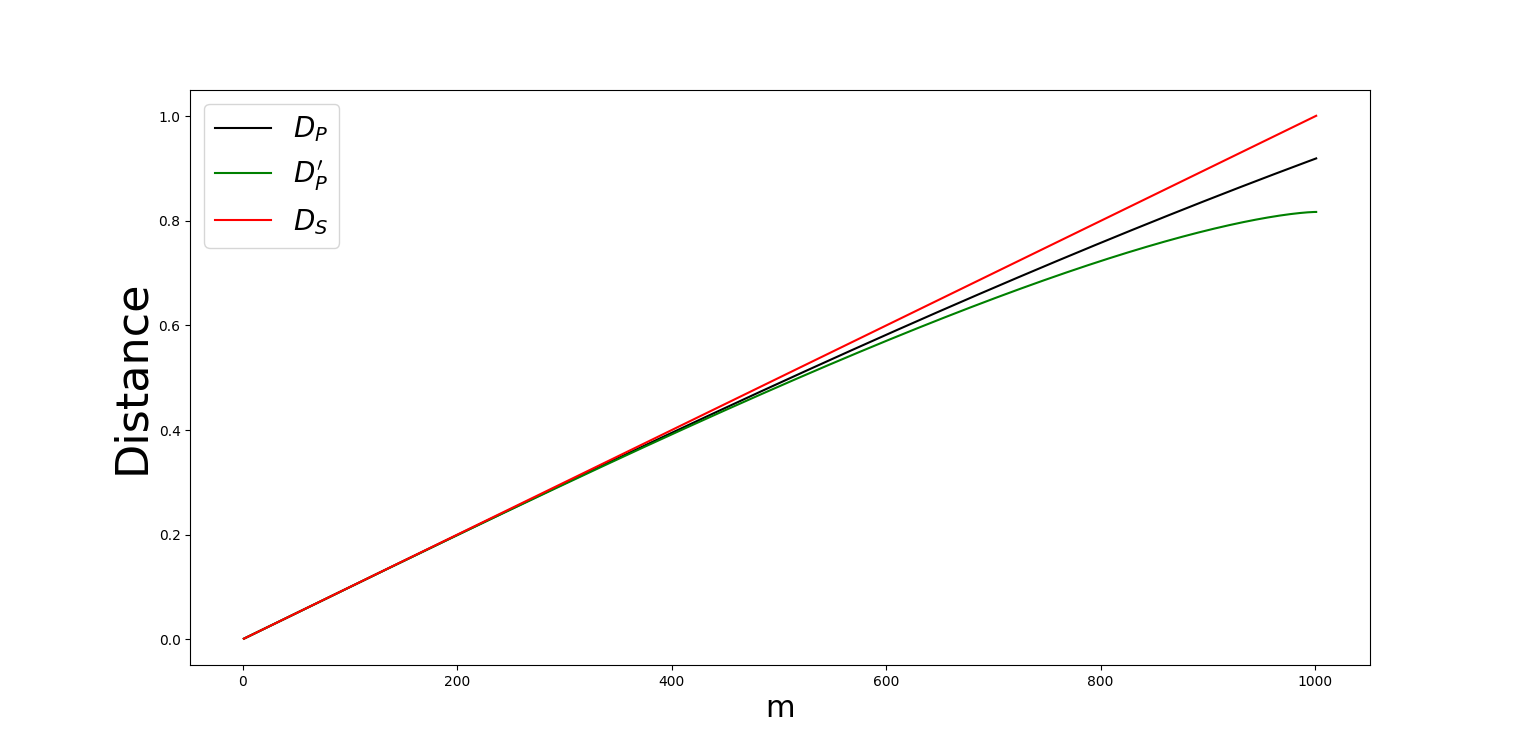}
  \caption{$\epsilon=0.001, m\leq 1001$}
  \label{fig:C}
 \end{subfigure}
 \hfill 
 \begin{subfigure}[b]{0.475\textwidth}
 \centering
 \includegraphics[width=7cm,height=5cm]{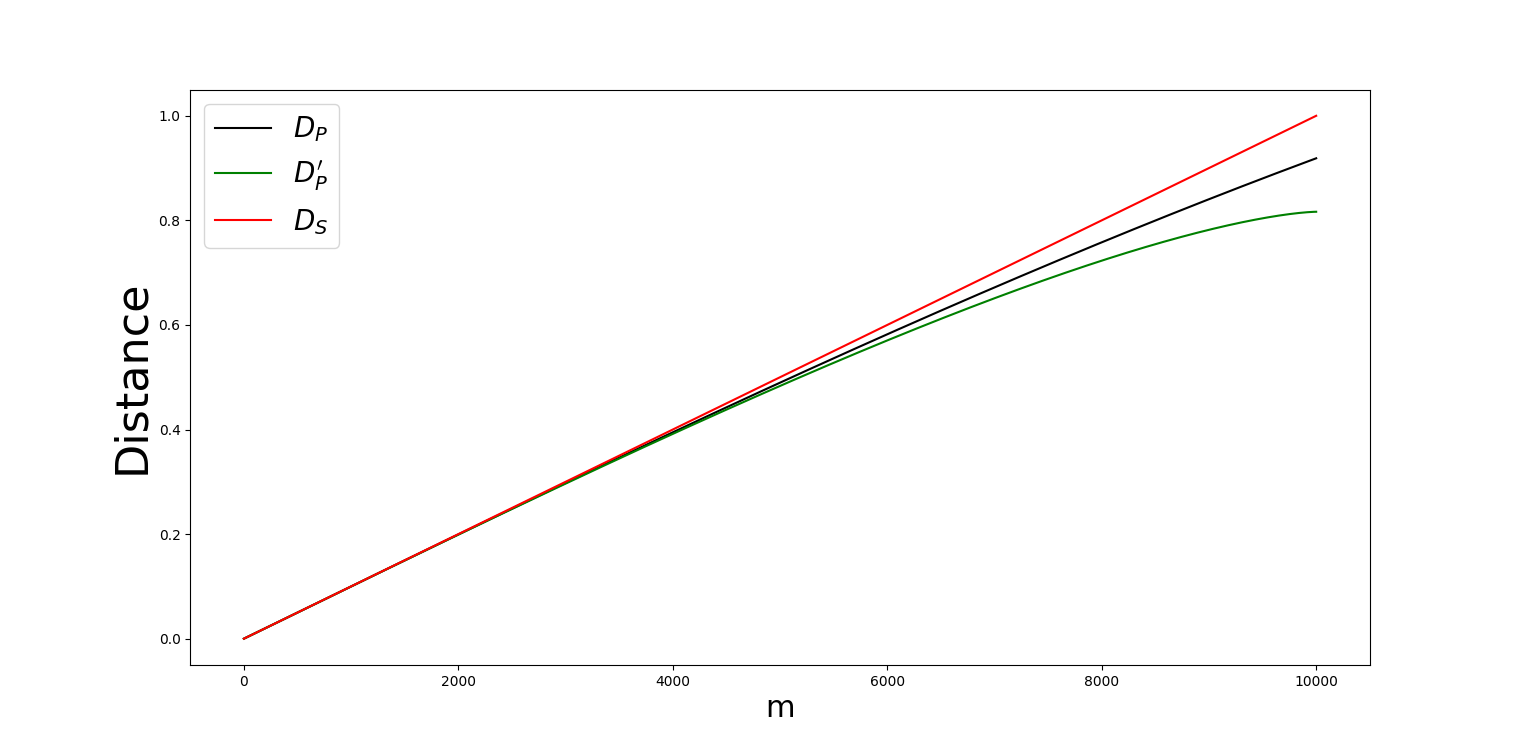}
  \caption{$\epsilon=0.0001, m\leq 10001$}
  \label{fig:D}
 \end{subfigure}
\caption{The error propagation for different values of $\epsilon$ and $m$. $\dph(U_i,V_i)\leq\epsilon$ for each $i$. We plot $m$ along X-axis. In the Y-axis the black line shows $\dph(\overline{U}_m,\overline{V}_m)$, the green line shows $\dph'(\overline{U}_m,\overline{V}_m)$ and the red line shows the sum-of-error $D_S=\sum_i\epsilon_i$. Here $\overline{U}_m=\prod_{i=m}^1U_i$ and $\overline{V}_m=\prod_{i=m}^1V_i$. $\dph'$ is the distance obtained using approximation-I.}
\label{fig:approx1}
\end{figure}
In Figure \ref{fig:approx1} we show the error propagation for different values of $\epsilon$ and $m$. Here $\epsilon$ is the upper bound on error for each unitary i.e. $\dph(U_i,V_i)\leq\epsilon$ for each $i$. The black line shows the error derived from Lemma \ref{lem:mult2}, the green line shows the error from Equation \ref{eqn:approx1} and the red line shows the sum-of-error bound. We see that the first two bounds are definitely less than the sum-of-error bound and the approximate bound derived in Equation \ref{eqn:approx1} closely follows the bound derived from Lemma \ref{lem:mult2}. We have shown the graphs for $\epsilon=10^{-1}, 10^{-2}, 10^{-3}, 10^{-4}$. Since the difference is more prominent for higher values of $\epsilon$, so we did not give the graphs for $\epsilon<10^{-4}$.

\subsection{Comparison with Approximation-II}
\label{app:approx2}

\begin{figure}
 \centering
 \begin{subfigure}[b]{0.475\textwidth}
 \centering
  \includegraphics[width=7cm,height=5cm]{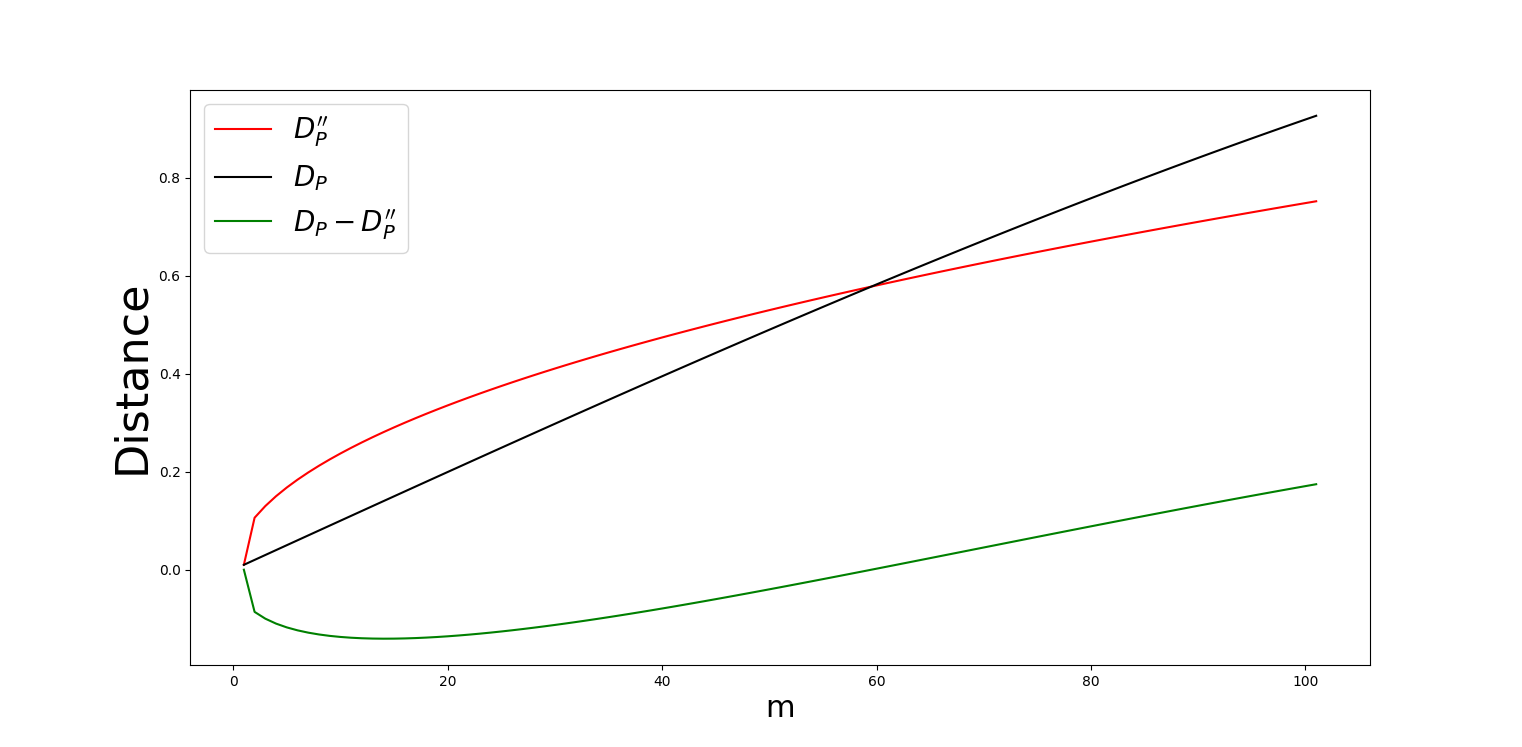}
  \caption{$\epsilon=10^{-2}, m\leq 101$}
  \label{fig:A}
 \end{subfigure}
 \hfill
 \begin{subfigure}[b]{0.475\textwidth}
 \centering
 \includegraphics[width=7cm,height=5cm]{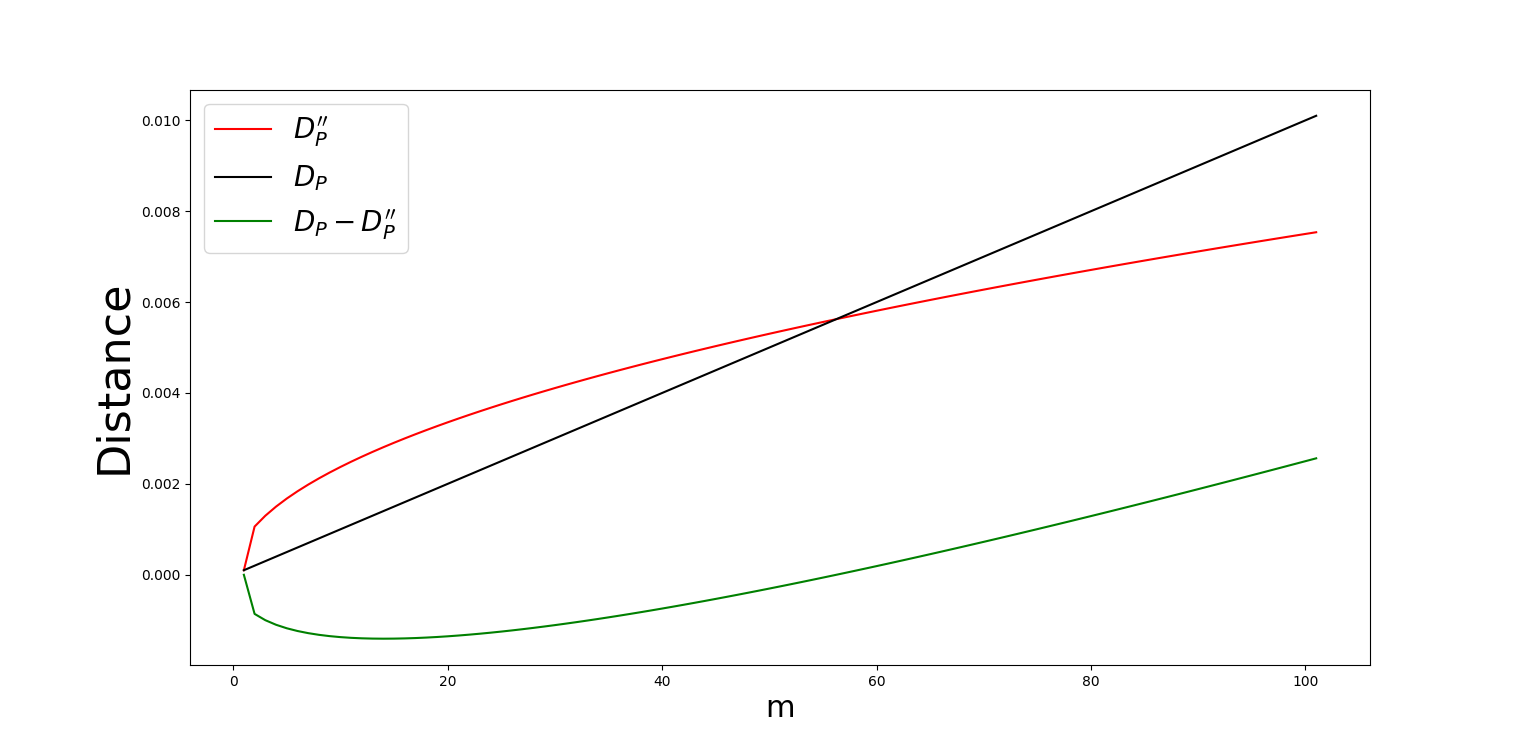}
  \caption{$\epsilon=10^{-4}, m\leq 101$}
  \label{fig:B}
 \end{subfigure}
 \vskip\baselineskip
 \begin{subfigure}[b]{0.475\textwidth}
 \centering
 \includegraphics[width=7cm,height=5cm]{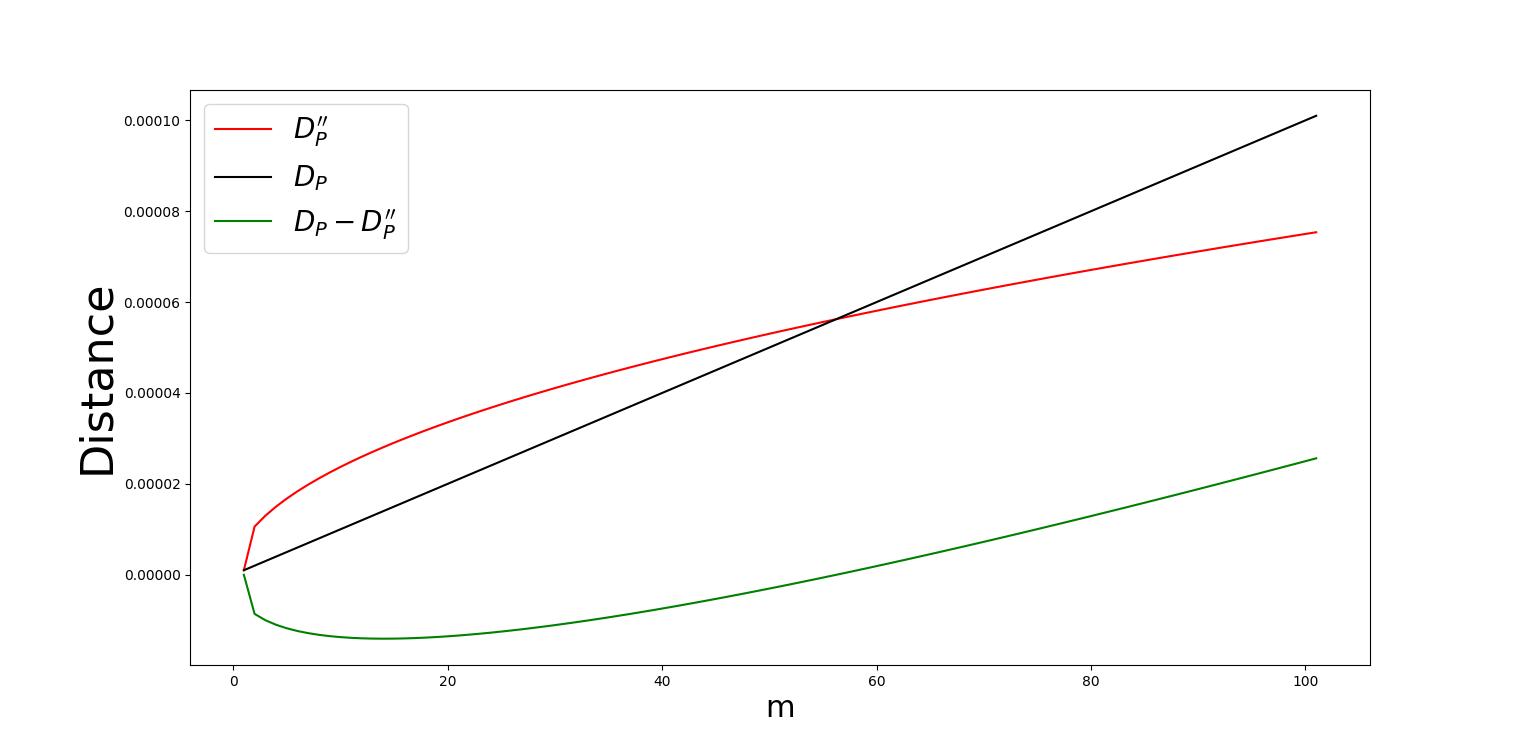}
  \caption{$\epsilon=10^{-6}, m\leq 101$}
  \label{fig:C}
 \end{subfigure}
 \hfill 
 \begin{subfigure}[b]{0.475\textwidth}
 \centering
 \includegraphics[width=7cm,height=5cm]{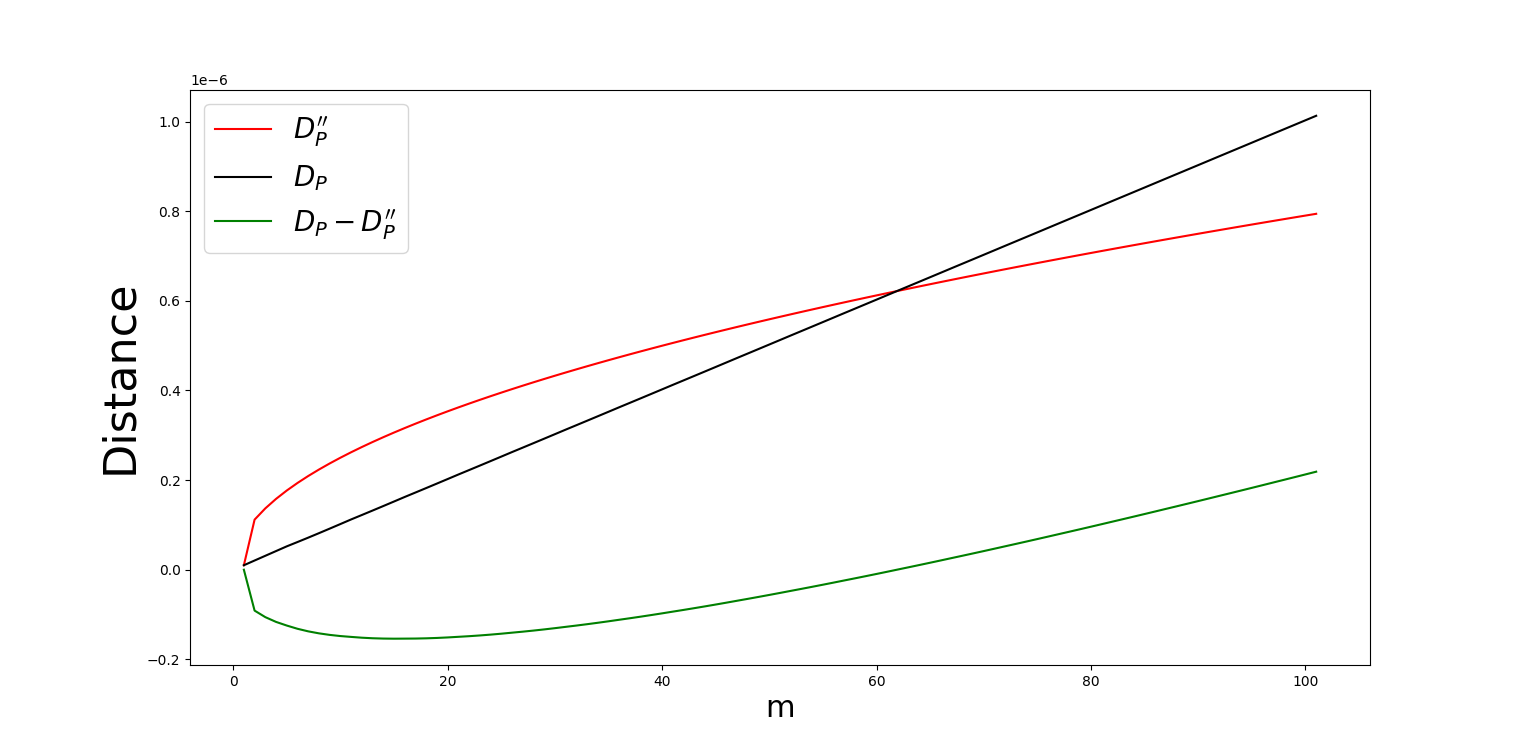}
  \caption{$\epsilon=10^{-8}, m\leq 101$. Y-axis scaled by $10^{-6}.$}
  \label{fig:D}
 \end{subfigure}
\caption{The error propagation for different values of $\epsilon$ and $m$. $\dph(U_i,V_i)\leq\epsilon$ for each $i$. We plot $m$ along X-axis. In the Y-axis the black line shows $\dph(\overline{U}_m,\overline{V}_m)$, the red line shows $\dph''(\overline{U}_m,\overline{V}_m)$ and the green line shows the difference $\dph-\dph''$. Here $\overline{U}_m=\prod_{i=m}^1U_i$ and $\overline{V}_m=\prod_{i=m}^1V_i$. $\dph''$ is the distance obtained using approximation-II.}
\label{fig:approx2}
\end{figure}

In Figure \ref{fig:approx2} we compare $\dph(\overline{U}_m,\overline{V}_m)$, the bound derived from Lemma \ref{lem:mult2} (black line) and the one derived in Equation \ref{eqn:approx2} (red line). The green line shows the difference. We take $c=7.5$ and $m\leq 100$. For even lower values of $\epsilon$ i.e. $< 10^{-8}$ the difference is even less.

\section{Error accumulation for tensor product of unitaries}
\label{app:tensor}

\begin{figure}
 \centering
 \begin{subfigure}[b]{0.475\textwidth}
 \centering
  \includegraphics[width=7cm,height=5cm]{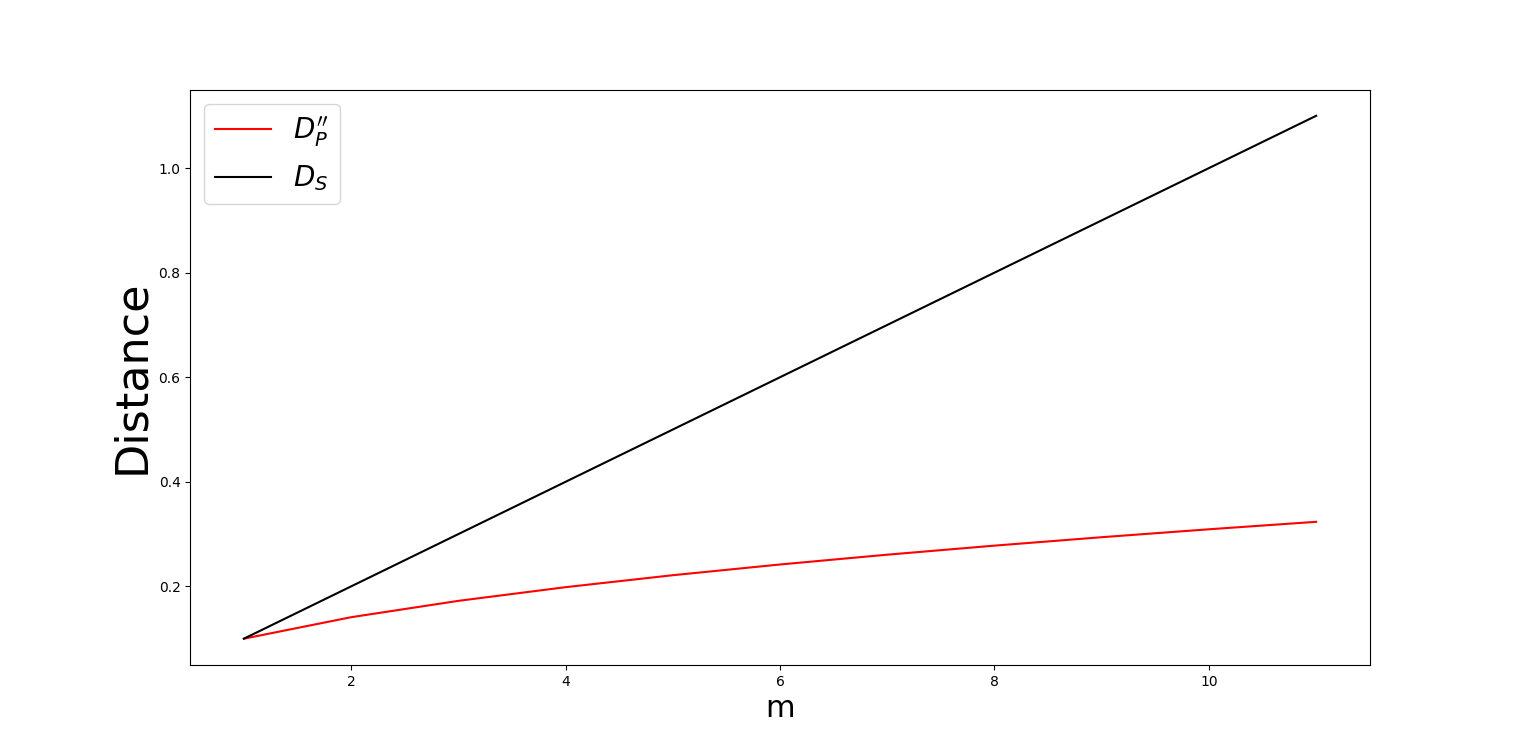}
  \caption{$\epsilon=0.1, m\leq 11$}
  \label{fig:A}
 \end{subfigure}
 \hfill
 \begin{subfigure}[b]{0.475\textwidth}
 \centering
 \includegraphics[width=7cm,height=5cm]{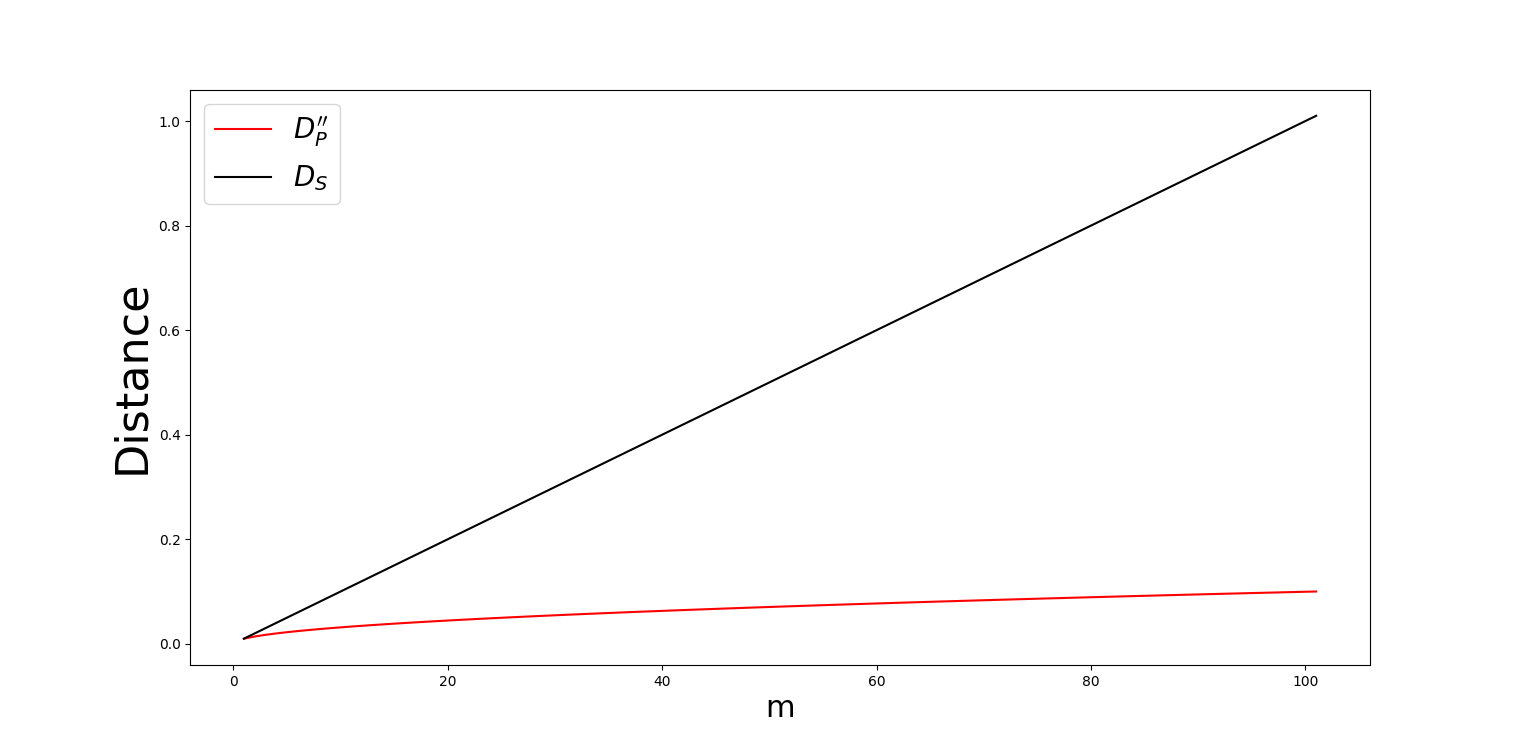}
  \caption{$\epsilon=0.01, m\leq 101$}
  \label{fig:B}
 \end{subfigure}
 \vskip\baselineskip
 \begin{subfigure}[b]{0.475\textwidth}
 \centering
 \includegraphics[width=7cm,height=5cm]{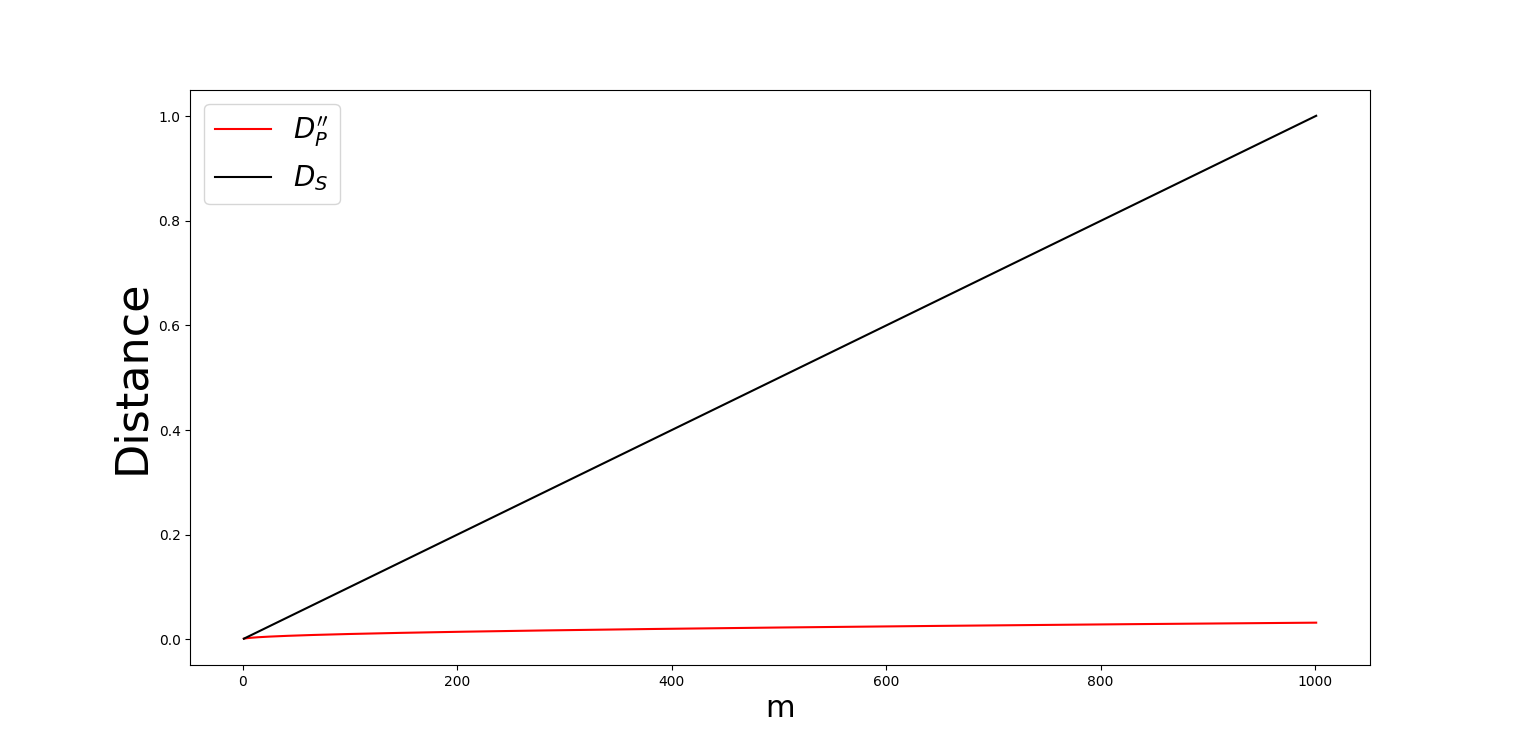}
  \caption{$\epsilon=0.001, m\leq 1001$}
  \label{fig:C}
 \end{subfigure}
 \hfill 
 \begin{subfigure}[b]{0.475\textwidth}
 \centering
 \includegraphics[width=7cm,height=5cm]{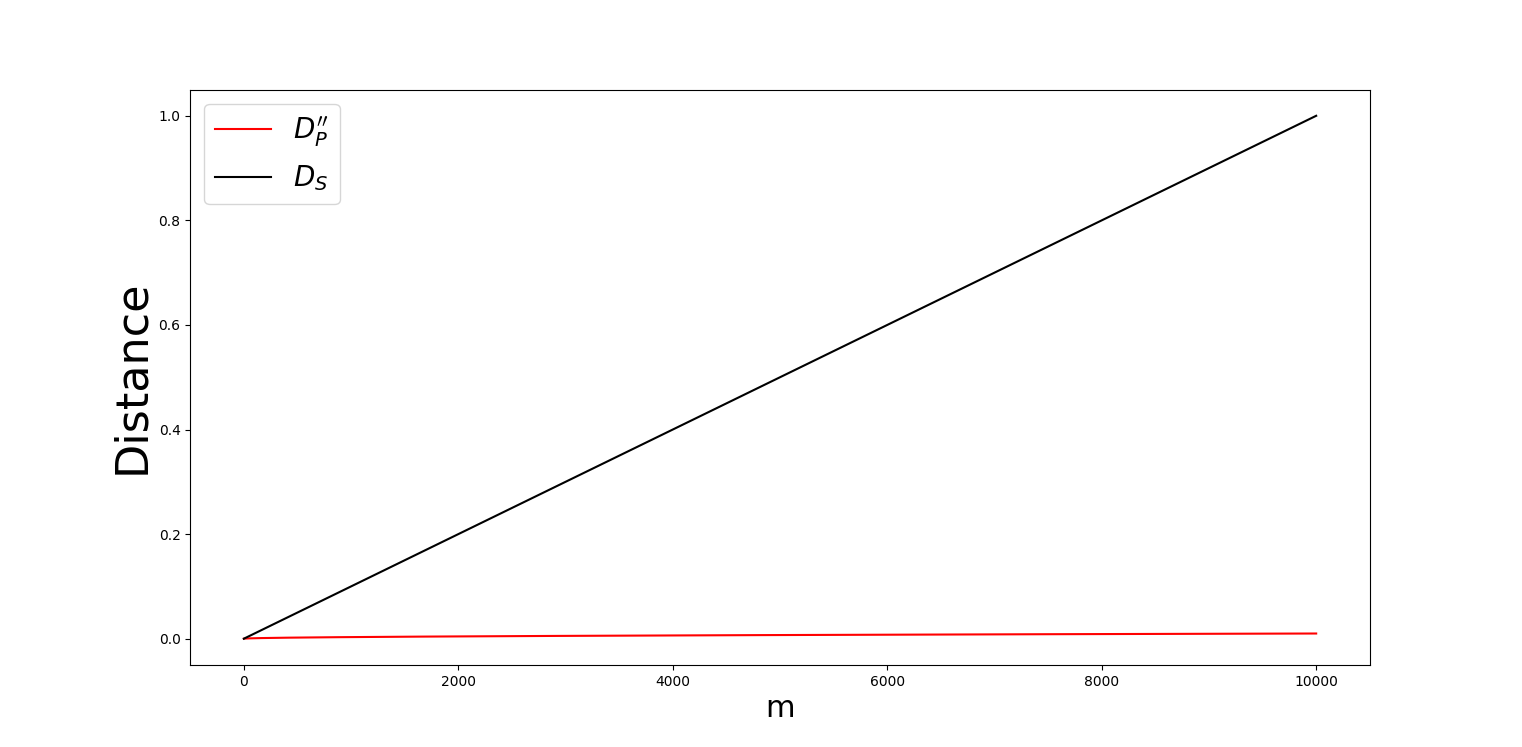}
  \caption{$\epsilon=0.0001, m\leq 10001$}
  \label{fig:D}
 \end{subfigure}
\caption{The error propagation for different values of $\epsilon$ and $m$. $\dph(U_i,V_i)\leq\epsilon$ for each $i$. We plot $m$ along X-axis. In the Y-axis the black line shows the sum-of-error and the red line shows $\dph(\bigotimes_{i=1}^mU_i,\bigotimes_{i=1}^m,V_i)$.}
\label{fig:tensor}
\end{figure}

In Figure \ref{fig:tensor} we show the error growth when the unitaries are in tensor product. That is, we use the bound in Lemma \ref{lem:composeTensor} (Section \ref{compose:tensor}) and compare with the sum-of-error bound.

\end{document}